\documentclass{elsart}

\usepackage{axodraw}
\usepackage{epsfig}
\usepackage{color}

\newcommand{\be}{\begin{equation}}
\newcommand{\ee}{\end{equation}}
\newcommand{\bea}{\begin{eqnarray}}
\newcommand{\eea}{\end{eqnarray}}
\newcommand{\nn}{\nonumber \\}
\newcommand{\as}{\alpha_s}
\newcommand{\ep}{\varepsilon}
\newcommand{\msb}{\overline{\mbox{MS}}}
\newcommand{\mtbar}{\overline{m_t}}

\def\phgt{50}   % all picture height
\def\pwc{100}   %     picture width

\def\phgth{25}   % half picture height
\newcommand{\masterpicture}[1]{\;\parbox[t]{\pwc pt}{
\begin{picture}(\pwc,\phgt)(0,\phgth)
#1
\SetScale{1}
\end{picture}}\;}

\begin{document}
  
  \begin{frontmatter}
    
    \title{\bf Single Scale Tadpoles and ${\mathcal O}(G_F m_t^2 \as^3)$
      Corrections to the $\rho$ Parameter}
    
    \author{R. Boughezal}
    \address{Institut f\"ur Theoretische Physik
      und Astrophysik, Universit\"at W\"urzburg, \\
      Am Hubland, D-97074 W\"urzburg, Germany}
    
    \author{M. Czakon}
    \address{Institut f\"ur Theoretische Physik
      und Astrophysik, Universit\"at W\"urzburg, \\
      Am Hubland, D-97074 W\"urzburg, Germany, \\
      Department of Field Theory and Particle Physics,
      Institute of Physics, \\
      University of Silesia, Uniwersytecka 4, PL-40007 Katowice,
      Poland}
    
    \begin{abstract}
      
      We present a new set of high precision numerical values of
      four-loop single-scale vacuum integrals, which we subsequently
      use to obtain the non-singlet corrections to the $\rho$
      parameter at ${\mathcal O}(G_F m_t^2 \as^3)$. Our result for
      $\Delta \rho$ is in agreement with the recent calculation
      \cite{Chetyrkin:2006bj}.
      
    \end{abstract}
    
  \end{frontmatter}
  
  \section{Introduction}
  
  Single-scale four-loop vacuum integrals have attracted a lot of
  attention in recent years. Even though the first applications were
  connected to the calculation of anomalous dimensions
  \cite{vanRitbergen:1997va,Czakon:2004bu}, the class of solved
  problems counts by now such topics as the pressure in hot QCD
  \cite{Kajantie:2002wa}, coupling constant and mass decoupling
  relations \cite{Schroder:2005hy,Chetyrkin:2005ia}, moments of the
  hadronic production cross section
  \cite{Boughezal:2006px,Chetyrkin:2006xg}, and corrections to the
  $\rho$ parameter \cite{Schroder:2005db,Chetyrkin:2006bj}. Studies of
  four-loop tadpoles have also led to new ideas in computational
  techniques, such as the introduction of special integral bases 
  \cite{Chetyrkin:2006dh}.

  This impressive progress has been made possible to a large extent by
  the Laporta algorithm for the reduction of integrals to masters
  described in \cite{Laporta:2001dd}, and by the difference equation
  method for the numerical evaluation of the masters proposed in the
  same publication. The first sets of integrals have been evaluated
  precisely using these principles \cite{Laporta:2002pg,Schroder:2005va}.
  At present, other methods are also available, see \cite{Kniehl:2005yc}
  and \cite{Chetyrkin:2006bj}.

  One of the applications of four-loop tadpoles mentioned at the
  beginning concerns a quantity of primary importance in the area of
  electroweak physics, namely the $\rho$ parameter introduced by
  Veltman \cite{Veltman:1977kh}. Defined as the ratio of the charged
  and neutral current strengths, it differs from its leading-order
  value of one, by a shift which can be expressed through the
  transverse parts of $W$ and $Z$ boson self-energies

  \be
  \Delta \rho = \frac{\Pi_Z(0)}{M_Z^2}-\frac{\Pi_W(0)}{M_W^2}.
  \ee

  This shift occurs as a universal correction in all electroweak
  observables and is thus related to the indirect prediction of the
  Higgs boson mass from the experimental data, and in particular from
  the $W$ boson mass \cite{Awramik:2003rn} and the effective weak
  mixing angle \cite{Awramik:2004ge}.

  In view of the importance of $\Delta \rho$, several corrections have
  been computed. In particular, the two- \cite{QCD2L} and three-loop
  \cite{QCD3L} QCD effects, and various electroweak effects in the
  limit of a large top quark mass \cite{largetop} have been accounted
  for. At the three-loop level, the leading behavior in the
  limit of a large Higgs boson mass is also available
  \cite{radja}. From now on, we will denote the QCD corrections to
  $\Delta \rho$ in leading order in the electroweak interaction by
  $\delta \rho$.

  At the four-loop level, the singlet QCD corrections, {\it i.e.}
  corrections where the external gauge bosons couple to different
  fermion loops have been evaluated in
  \cite{Schroder:2005db}. Motivated by that publication, we started the
  calculation of the non-singlet contributions, which are obtained by
  attaching gluons (with possible fermion loop insertions) to the
  leading one-loop diagrams. The major obstacle to overcome is the
  calculation of the many new master integrals. It is the purpose of
  the present work to present our results for those integrals and
  apply them to the calculation of the four-loop non-singlet QCD
  corrections to $\Delta \rho$.

  Recently, Ref.~\cite{Chetyrkin:2006bj} containing a result
  for the very same corrections appeared. Anticipating the content of
  Section~\ref{results}, we can state that we agree with this
  calculation.
  
  This paper is organized as follows. In the next section we present
  high precision numeric expansions of the master integrals. We then
  give our on-shell result and conclusions. An appendix contains the
  corrections expressed in the $\msb$ scheme.

  \section{Master Integrals}

  \label{masters}

  Upon reducing the complete set of scalar integrals occurring in the
  diagrams contributing to $\Delta \rho$ at the four-loop level, we
  are left with 65 masters. However, the latter number is only correct
  if we consider the integration-by-parts identities for a given
  prototype and not those of its parents\footnote{integrals with a
  larger number of lines, such that removing some lines and merging
  the remaining vertices leads to the original integral.}. It has been
  noticed in the case of two \cite{Czakon:2006pf} and three-loop
  \cite{Laporta:1997zy} on-shell propagators that the identities of
  the parents can further reduce the set of master
  integrals. Interestingly, the same happens also in the present 
  calculation. In fact, we found the following two relations

  \epsfig{file=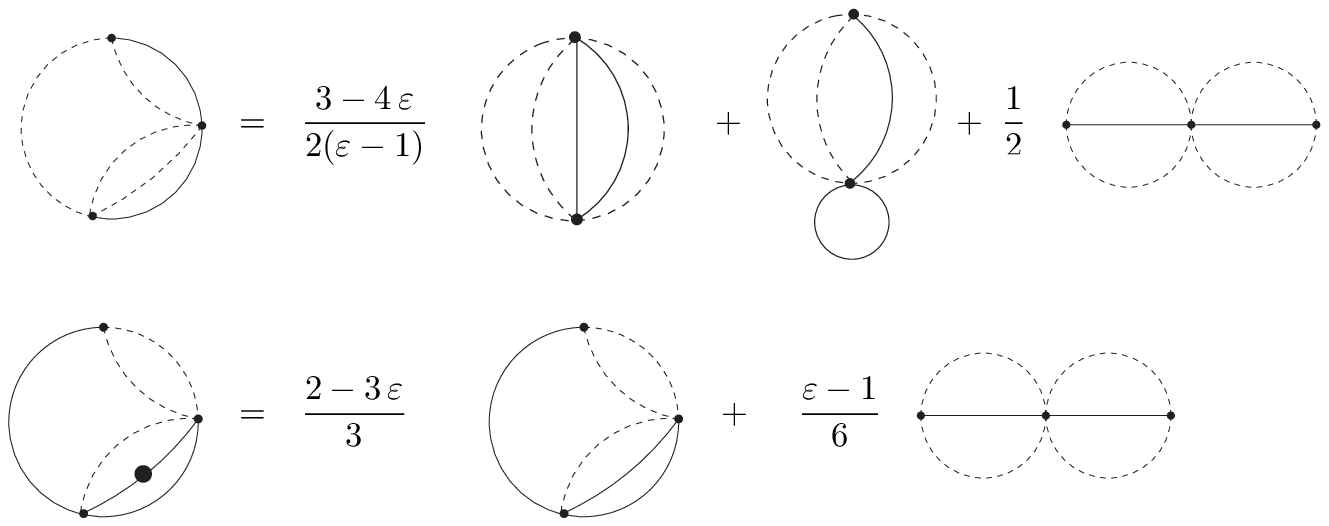}

  where the continuous and dashed lines denote massive and massless
  lines respectively.  The final number of master integrals is,
  therefore, 63. Of these 23 have either been already presented in the
  literature \cite{Schroder:2005va}, or are products of vacuum
  integrals with a lower number of loops, or can be trivially
  expressed in terms of gamma functions. Ultimately, we need only to
  calculate 40 new integrals. At this point, a few words are due to
  explain our choice of masters. Since the algorithm we use for
  numerical evaluation can provide high precision numbers and deep
  expansions in the dimensional regularization parameter
  $\ep$\footnote{we assume the dimension of space-time to be
  $d=4-2\ep$.}, we let the reduction software chose the masters
  automatically, keeping, however, integrals with dots (higher powers
  of denominators) instead of irreducible numerators.

  As mentioned in the Introduction, the method of difference equations
  \cite{Laporta:2001dd} provides a particularly efficient approach for
  the evaluation of vacuum integrals. Its main steps are
  \begin{enumerate}
    \item introduction of a symbolic power, $x$, on a chosen massive line
      of a given vacuum integral, which is then a function $T(x,\ep)$;
    \item determination of a difference equation in $x$,
    \[ \sum_{i=0}^{n} c_i(x,\ep)T(x+i,\ep) = U(x,\ep), \]
    where $c_i(x,\ep)$ are rational functions and $U(x,\ep)$ is given by
    vacuum integrals with less lines;
    \item determination of the boundary condition at $x\rightarrow
    \infty$, which is equivalent to finding the low momentum expansion
    of the propagator subloop, when the line carrying power $x$ is cut;
    \item solution of the difference equation after expansion in
    $\ep$, assuming that every term of the expansion of $T(x,\ep)$
    takes the form of a factorial series 
    \[ \mu^x \sum_{s=0}^{\infty} 
    \frac{a_s \Gamma(x+1)}{\Gamma(x-K+s+1)}, \]
    where $\mu$, $K$ and $a_s$ remain to be determined from the
    difference equation and the boundary condition.
  \end{enumerate}
  Further details can be found in the original work
  \cite{Laporta:2001dd}. We note here only that the difference
  equations are derived from integration-by-parts identities, which
  are now more complicated, because they involve two variables and not
  just one, as those of the reduction to masters. The procedure we
  sketched above is recursive and performed numerically. The recursion
  leads to the loss of many terms in the expansion in $\ep$.  Even if
  we start with a three-loop input up to $\ep^{10}$, in some cases we are
  only left with terms up to $\ep^2$ in the four-loop  results. Apart
  from these exceptions, all the 5-, 6- and  7-liners are provided up
  to $\ep^4$. The remaining 8-, and 9-liners are given up to $\ep^3$.
  The depth of the expansion of our four-loop masters would be sufficient
  for five-loop calculations. As far as the number of digits is
  concerned, our goal was to obtain at least 30 digits for all  the
  masters. This has turned out to be difficult for the last three
  9-liners shown bellow, because of weak convergence. In the worst
  case, we have  still obtained 18 digits.

  Before we list our results, we have to define the normalization of
  the integrals. We simply assume that the integration measure is
  $(e^{\ep \gamma}/i\pi^{d/2})^4 \int d^d k_1...d^d k_4$ and that the
  propagators in the subsequent expressions are
  $1/(k^2-m^2)$ and we set everywhere the mass to unity. The values of
  the computed integrals are

  \bea
  %
  % PR61[1, 1, 1, 1, 1, 0, 0, 0, 0, 0]
  \masterpicture{
    %\begin{center}
    \fcolorbox{white}{white}{
      \SetScale{0.39}
      \SetOffset(-48,32)
      %  \begin{picture}(151,151) (211,-89)
      \SetWidth{0.5}
      \SetColor{Black}
      \CArc(287,-14)(67.36,108,468)
      \Line(286,53)(286,-82)
      \DashCArc(345.82,-14)(89.82,131.76,228.24){4}
      \DashCArc(224.03,-13.04)(91.97,-47.64,46.79){4}
      \Vertex(286,53){2.83}
      \Vertex(286,-81){2.83}
      %  \end{picture}
    }
    %\end{center}
  } &=&{} 
  + 0.250000000000000000000000000000 \; \ep^{-4}
  \nn&&{} + 1.00000000000000000000000000000 \; \ep^{-3}
  \nn&&{} + 2.84330036675744655156954091666 \; \ep^{-2}
  \nn&&{} + 5.78154361042103255558936205723 \; \ep^{-1}
  \nn&&{} + 22.9556218817054224703296183863
  \nn&&{} + 80.8955061678534024741104898218 \; \ep
  \nn&&{} + 1085.28365870727983857702515746 \; \ep^{2}
  \nn&&{} + 4545.30388413442580391360145658 \; \ep^{3}
  \nn&&{} + 35998.9938326326556313329606141 \; \ep^{4}
  \\
  %
  % PR61[1, 1, 1, 1, 2, 0, 0, 0, 0, 0]
  \masterpicture{
    %\begin{center}
    \fcolorbox{white}{white}{
      \SetScale{0.39}
      \SetOffset(-48,32)
      %  \begin{picture}(151,151) (211,-89)
      \SetWidth{0.5}
      \SetColor{Black}
      \CArc(287,-14)(67.36,108,468)
      \Line(286,53)(286,-82)
      \DashCArc(345.82,-14)(89.82,131.76,228.24){4}
      \DashCArc(224.03,-13.04)(91.97,-47.64,46.79){4}
      \Vertex(286,53){2.83}
      \Vertex(286,-81){2.83}
      \Vertex(256,-13){5.66}
      %  \end{picture}
    }
    %\end{center}
  } &=&{} 
  - 0.250000000000000000000000000000 \; \ep^{-4}
  \nn&&{} - 1.12500000000000000000000000000 \; \ep^{-3}
  \nn&&{} - 3.32246703342411321823620758332 \; \ep^{-2}
  \nn&&{} - 6.69920730307326055744986988775 \; \ep^{-1}
  \nn&&{} + 30.0253835218317623192659472127
  \nn&&{} + 254.111031964704093539375826207 \; \ep
  \nn&&{} + 1805.36593366425290012409359932 \; \ep^{2}
  \nn&&{} + 8898.24672362282959032538518673 \; \ep^{3}
  \nn&&{} + 43751.2551818001444717625115350 \; \ep^{4}
  \\
  %
  % PR65[1, 1, 1, 1, 1, 1, 0, 0, 0, 0]
  \masterpicture{
    %\begin{center}
    \fcolorbox{white}{white}{
      \SetScale{0.44}
      \SetOffset(-78,55)
      %  \begin{picture}(129,133) (252,-131)
      \SetWidth{0.5}
      \SetColor{Black}
      \DashCArc(320,-64)(62,90,102){4}                    % line 1
      \DashCArc(320,-64)(62,-12,90){4}                    % line 2
      \DashCArc(320,-64)(62,102,348){4}                   % line 3
      \CArc(225,13)(95,-70,-9)                     % line 4
      \CArc(415,13)(95,-171,-110)                  % line 5
      \DashCArc(320,-0.66)(98.35,-128,-52){4}             % line 6
      \Vertex(320,-2){2.83}
      \Vertex(260,-77){2.83}
      \Vertex(380,-77){2.83}
      %  \end{picture}
    }
    %\end{center}
  } &=&{} 
  + 0.0833333333333333333333333333333 \; \ep^{-4}
  \nn&&{} + 0.416666666666666666666666666667 \; \ep^{-3}
  \nn&&{} + 0.857489011141371072745402527774 \; \ep^{-2}
  \nn&&{} - 3.38659426575776059720149630689 \; \ep^{-1}
  \nn&&{} - 74.0498948425972614880435957857
  \nn&&{} - 471.009207065575912327978772140 \; \ep
  \nn&&{} - 3088.55385540997021022612585453 \; \ep^{2}
  \nn&&{} - 14498.1092918745193546816622992 \; \ep^{3}
  \nn&&{} - 75204.1336314891447627178739121 \; \ep^{4}
  \\
  %
  % PR98[1, 1, 1, 1, 1, 1, 0, 0, 0, 0]
  \masterpicture{
    %\begin{center}
    \fcolorbox{white}{white}{
      \SetScale{0.44}
      \SetOffset(-78,55)
      %  \begin{picture}(129,133) (252,-131)
      \SetWidth{0.5}
      \SetColor{Black}
      \CArc(320,-64)(62,90,102)                    % line 1
      \DashCArc(320,-64)(62,-12,90){4}                    % line 2
      \CArc(320,-64)(62,102,348)                   % line 3
      \DashCArc(225,13)(95,-70,-9){4}                     % line 4
      \DashCArc(415,13)(95,-171,-110){4}                  % line 5
      \CArc(320,-0.66)(98.35,-128,-52)             % line 6
      \Vertex(320,-2){2.83}
      \Vertex(260,-77){2.83}
      \Vertex(380,-77){2.83}
      %  \end{picture}
    }
    %\end{center}
  } &=&{} 
  + 0.333333333333333333333333333333 \; \ep^{-4}
  \nn&&{} + 1.91666666666666666666666666667 \; \ep^{-3}
  \nn&&{} + 6.51328937789881762431494344443 \; \ep^{-2}
  \nn&&{} + 14.5114757391222468152054282866 \; \ep^{-1}
  \nn&&{} - 59.1987779967727811953138849540
  \nn&&{} - 549.042726795028324060762834254 \; \ep
  \nn&&{} - 4293.52617296657045021413760744 \; \ep^{2}
  \nn&&{} - 20846.8297417686461476497501336 \; \ep^{3}
  \nn&&{} - 110899.174077030571795400409124 \; \ep^{4}
  \\
  %
  % PR98[1, 1, 1, 1, 1, 2, 0, 0, 0, 0]
  \masterpicture{
    %\begin{center}
    \fcolorbox{white}{white}{
      \SetScale{0.44}
      \SetOffset(-78,55)
      %  \begin{picture}(129,133) (252,-131)
      \SetWidth{0.5}
      \SetColor{Black}
      \CArc(320,-64)(62,90,102)                    % line 1
      \DashCArc(320,-64)(62,-12,90){4}                    % line 2
      \CArc(320,-64)(62,102,348)                   % line 3
      \DashCArc(225,13)(95,-70,-9){4}                     % line 4
      \DashCArc(415,13)(95,-171,-110){4}                  % line 5
      \CArc(320,-0.66)(98.35,-128,-52)             % line 6
      \Vertex(320,-2){2.83}
      \Vertex(260,-77){2.83}
      \Vertex(380,-77){2.83}
      \Vertex(320,-126){5.66}
      %  \end{picture}
    }
    %\end{center}
  } &=&{} 
  + 0.250000000000000000000000000000 \; \ep^{-4}
  \nn&&{} + 0.916666666666666666666666666667 \; \ep^{-3}
  \nn&&{} + 1.90580036675744655156954091666 \; \ep^{-2}
  \nn&&{} + 2.52073453954145394473436269644 \; \ep^{-1}
  \nn&&{} - 62.9263953267520790182858402690
  \nn&&{} - 258.692941139129800870656513172 \; \ep
  \nn&&{} - 2245.88328900424600912128374855 \; \ep^{2}
  \nn&&{} - 7643.28441760863244533368790482 \; \ep^{3}
  \nn&&{} - 48359.2665176408023251619751028 \; \ep^{4}
  \\
  %
  % PR132[1, 1, 1, 1, 1, 1, 0, 0, 0, 0]
  \masterpicture{
    %\begin{center}
    \fcolorbox{white}{white}{
      \SetScale{0.44}
      \SetOffset(-78,55)
      %  \begin{picture}(129,133) (252,-131)
      \SetWidth{0.5}
      \SetColor{Black}
      \CArc(320,-64)(62,90,102)                    % line 1
      \CArc(320,-64)(62,-12,90)                    % line 2
      \CArc(320,-64)(62,102,348)                   % line 3
      \DashCArc(225,13)(95,-70,-9){4}                     % line 4
      \DashCArc(415,13)(95,-171,-110){4}                  % line 5
      \CArc(320,-0.66)(98.35,-128,-52)             % line 6
      \Vertex(320,-2){2.83}
      \Vertex(260,-77){2.83}
      \Vertex(380,-77){2.83}
      %  \end{picture}
    }
    %\end{center}
  } &=&{} 
  + 0.583333333333333333333333333333 \; \ep^{-4}
  \nn&&{} + 3.58333333333333333333333333333 \; \ep^{-3}
  \nn&&{} + 13.6690897446562641758844843611 \; \ep^{-2}
  \nn&&{} + 38.3870766266324744689703482793 \; \ep^{-1}
  \nn&&{} - 51.2252212801189152418819912275
  \nn&&{} - 477.947460995194619277285562192 \; \ep
  \nn&&{} - 6127.93433249128299350151222877 \; \ep^{2}
  \nn&&{} - 23752.7464620411276168084881607 \; \ep^{3}
  \nn&&{} - 161921.400195144895866566378264 \; \ep^{4}
  \\
  %
  % PR132[1, 1, 1, 1, 1, 2, 0, 0, 0, 0]
  \masterpicture{
    %\begin{center}
    \fcolorbox{white}{white}{
      \SetScale{0.42}
      \SetOffset(-70,55)
      %  \begin{picture}(129,133) (252,-131)
      \SetWidth{0.5}
      \SetColor{Black}
      \CArc(320,-64)(62,90,102)                    % line 1
      \CArc(320,-64)(62,-12,90)                    % line 2
      \CArc(320,-64)(62,102,348)                   % line 3
      \DashCArc(225,13)(95,-70,-9){4}                     % line 4
      \DashCArc(415,13)(95,-171,-110){4}                  % line 5
      \CArc(320,-0.66)(98.35,-128,-52)             % line 6
      \Vertex(320,-2){2.83}
      \Vertex(260,-77){2.83}
      \Vertex(380,-77){2.83}
      \Vertex(320,-126){5.66}
      %  \end{picture}
    }
    %\end{center}
  } &=&{} 
  + 0.333333333333333333333333333333 \; \ep^{-4}
  \nn&&{} + 1.33333333333333333333333333333 \; \ep^{-3}
  \nn&&{} + 3.42995604456548429098161011110 \; \ep^{-2}
  \nn&&{} + 5.55063171514827193110902146145 \; \ep^{-1}
  \nn&&{} - 57.5851192392905729834630220283
  \nn&&{} - 226.906688638637087111211732760 \; \ep
  \nn&&{} - 2357.42884823258106273364123809 \; \ep^{2}
  \nn&&{} - 6885.09836097771127634712790473 \; \ep^{3}
  \nn&&{} - 51888.4446361973078707935677603 \; \ep^{4}
  \\
  %
  % PR100[1, 1, 1, 1, 1, 1, 0, 0, 0, 0]
  \masterpicture{
    %\begin{center}
    \fcolorbox{white}{white}{
      \SetScale{0.44}
      \SetOffset(-78,55)
      %  \begin{picture}(129,133) (252,-131)
      \SetWidth{0.5}
      \SetColor{Black}
      \CArc(320,-64)(62,90,102)                    % line 1
      \CArc(320,-64)(62,-12,90)                    % line 2
      \CArc(320,-64)(62,102,348)                   % line 3
      \CArc(225,13)(95,-70,-9)                     % line 4
      \CArc(415,13)(95,-171,-110)                  % line 5
      \DashCArc(320,-0.66)(98.35,-128,-52){4}             % line 6
      \Vertex(320,-2){2.83}
      \Vertex(260,-77){2.83}
      \Vertex(380,-77){2.83}
      %  \end{picture}
    }
    %\end{center}
  } &=&{} 
  + 1.00000000000000000000000000000 \; \ep^{-4}
  \nn&&{} + 6.25000000000000000000000000000 \; \ep^{-3}
  \nn&&{} + 24.7065348003631195396114970000 \; \ep^{-2}
  \nn&&{} + 63.1857353991734145122797593446 \; \ep^{-1}
  \nn&&{} + 13.5916504314381310784889154404
  \nn&&{} - 595.072765215957129399869660562 \; \ep
  \nn&&{} - 6875.84113744124081014720730187 \; \ep^{2}
  \nn&&{} - 31308.1849304592520703889751908 \; \ep^{3}
  \nn&&{} - 192167.397958807277872297783882 \; \ep^{4}
  \\
  %
  % PR100[1, 1, 1, 1, 1, 2, 0, 0, 0, 0]
  \masterpicture{
    %\begin{center}
    \fcolorbox{white}{white}{
      \SetScale{0.44}
      \SetOffset(-78,55)
      %  \begin{picture}(129,133) (252,-131)
      \SetWidth{0.5}
      \SetColor{Black}
      \CArc(320,-64)(62,90,102)                    % line 1
      \CArc(320,-64)(62,-12,90)                    % line 2
      \CArc(320,-64)(62,102,348)                   % line 3
      \CArc(225,13)(95,-70,-9)                     % line 4
      \CArc(415,13)(95,-171,-110)                  % line 5
      \DashCArc(320,-0.66)(98.35,-128,-52){4}             % line 6
      \Vertex(320,-2){2.83}
      \Vertex(260,-77){2.83}
      \Vertex(380,-77){2.83}
      \Vertex(272,-24){5.66}
      %  \end{picture}
    }
    %\end{center}
  } &=&{} 
  + 0.416666666666666666666666666667 \; \ep^{-4}
  \nn&&{} + 1.75000000000000000000000000000 \; \ep^{-3}
  \nn&&{} + 4.95411172237352203039367930554 \; \ep^{-2}
  \nn&&{} + 4.75429224977896467341162845720 \; \ep^{-1}
  \nn&&{} - 37.4574649312968260550782554633
  \nn&&{} - 296.071923695141317174869536815 \; \ep
  \nn&&{} - 2022.20010652903000709521858304 \; \ep^{2}
  \nn&&{} - 8244.01633827202034423181172211 \; \ep^{3}
  \nn&&{} - 46235.4396348606702002905412792 \; \ep^{4}
  \\
  %
  % PR58[1, 1, 1, 1, 1, 1, 0, 0, 0, 0]
  \masterpicture{
    %\begin{center}
    \fcolorbox{white}{white}{
      \SetScale{0.44}
      \SetOffset(-21,42)
      %  \begin{picture}(129,129) (120,-101)
      \SetWidth{0.5}
      \SetColor{Black}
      \DashCArc(184,-38)(62,0,90){4}                      % line 1
      \CArc(184,-38)(62,90,258)                    % line 2
      \CArc(184,-38)(62,258,360)                   % line 3
      \DashCArc(258.11,38.74)(75.57,-168.75,-98.46){4}    % line 4
      \DashCArc(239.52,-106.92)(71.22,84.78,171.99){4}    % line 5
      \DashCArc(91.21,76.16)(191.15,-65.66,-35.93){4}     % line 6
      \Vertex(184,24){2.83}
      \Vertex(246,-36){2.83}
      \Vertex(171,-98){2.83}
      %  \end{picture}
    }
    %\end{center}
  } &=&{} 
  + 0.208333333333333333333333333333 \; \ep^{-4}
  \nn&&{} + 1.14583333333333333333333333333 \; \ep^{-3}
  \nn&&{} + 4.95577289461087423343304723609 \; \ep^{-2}
  \nn&&{} + 15.6523028485501530451923763258 \; \ep^{-1}
  \nn&&{} + 11.9062943058191243261235775944
  \nn&&{} - 71.7337613396528954521319262962 \; \ep
  \nn&&{} - 1308.40045975490723588094289795 \; \ep^{2}
  \nn&&{} - 6297.44410007372169525480920361 \; \ep^{3}
  \nn&&{} - 39381.3291846386139130702622754 \; \ep^{4}
  \\
  %
  % PR102[1, 1, 1, 1, 1, 1, 0, 0, 0, 0]
  \masterpicture{
    %\begin{center}
    %\fcolorbox{white}{white}{
    \SetScale{0.44}
    \SetOffset(-11,42)
    %  \begin{picture}(129,129) (120,-101)
    \SetWidth{0.5}
    \SetColor{Black}
    \DashCArc(184,-38)(62,0,90){4}                      % line 1
    \CArc(184,-38)(62,90,258)                    % line 2
    \CArc(184,-38)(62,258,360)                   % line 3
    \DashCArc(258.11,38.74)(75.57,-168.75,-98.46){4}    % line 4
    \DashCArc(239.52,-106.92)(71.22,84.78,171.99){4}    % line 5
    \CArc(91.21,76.16)(191.15,-65.66,-35.93)     % line 6
    \Vertex(184,24){2.83}
    \Vertex(246,-36){2.83}
    \Vertex(171,-98){2.83}
    %  \end{picture}
    %}
    %\end{center}
  } &=&{} 
  + 0.500000000000000000000000000000 \; \ep^{-4}
  \nn&&{} + 2.93750000000000000000000000000 \; \ep^{-3}
  \nn&&{} + 13.1961181336964528729448303333 \; \ep^{-2}
  \nn&&{} + 42.9395966422292576855124445264 \; \ep^{-1}
  \nn&&{} + 83.1358882000807808465322734823
  \nn&&{} + 27.1740919045453848408661874006 \; \ep
  \nn&&{} - 1554.56495046662560630454485042 \; \ep^{2}
  \nn&&{} - 9710.65466554205477699849510015 \; \ep^{3}
  \nn&&{} - 59651.1029754541134802977172410 \; \ep^{4}
  \\
  %
  % PR73[1, 1, 1, 1, 1, 1, 0, 0, 0, 0]
  \masterpicture{
    %\begin{center}
    \fcolorbox{white}{white}{
      \SetScale{0.44}
      \SetOffset(-21,42)
      %  \begin{picture}(129,129) (120,-101)
      \SetWidth{0.5}
      \SetColor{Black}
      \DashCArc(184,-38)(62,0,90){4}                      % line 1
      \CArc(184,-38)(62,90,258)                    % line 2
      \CArc(184,-38)(62,258,360)                   % line 3
      \DashCArc(258.11,38.74)(75.57,-168.75,-98.46){4}    % line 4
      \CArc(239.52,-106.92)(71.22,84.78,171.99)    % line 5
      \CArc(91.21,76.16)(191.15,-65.66,-35.93)     % line 6
      \Vertex(184,24){2.83}
      \Vertex(246,-36){2.83}
      \Vertex(171,-98){2.83}
      %  \end{picture}
    }
    %\end{center}
  } &=&{} 
  + 0.875000000000000000000000000000 \; \ep^{-4}
  \nn&&{} + 5.31250000000000000000000000000 \; \ep^{-3}
  \nn&&{} + 24.1272857172567359185353492916 \; \ep^{-2}
  \nn&&{} + 71.5869874700921357789204087564 \; \ep^{-1}
  \nn&&{} + 197.558371935133443514418819015
  \nn&&{} + 6.41187248372530806306191468163 \; \ep
  \nn&&{} - 1028.33128833387533834458667353 \; \ep^{2}
  \nn&&{} - 16249.8777552641957747492917792 \; \ep^{3}
  \nn&&{} - 65696.8989733875039196986140933 \; \ep^{4}
  \\
  %
  % PR85[1, 1, 1, 1, 1, 1, 0, 0, 0, 0]
  \masterpicture{
    %\begin{center}
    \fcolorbox{white}{white}{
      \SetScale{0.44}
      \SetOffset(-21,42)
      %  \begin{picture}(129,129) (120,-101)
      \SetWidth{0.5}
      \SetColor{Black}
      \CArc(184,-38)(62,0,90)                      % line 1
      \DashCArc(184,-38)(62,90,258){4}                    % line 2
      \CArc(184,-38)(62,258,360)                   % line 3
      \DashCArc(258.11,38.74)(75.57,-168.75,-98.46){4}    % line 4
      \CArc(239.52,-106.92)(71.22,84.78,171.99)    % line 5
      \CArc(91.21,76.16)(191.15,-65.66,-35.93)     % line 6
      \Vertex(184,24){2.83}
      \Vertex(246,-36){2.83}
      \Vertex(171,-98){2.83}
      %  \end{picture}
    }
    %\end{center}
  } &=&{} 
  + 0.625000000000000000000000000000 \; \ep^{-4}
  \nn&&{} + 4.12500000000000000000000000000 \; \ep^{-3}
  \nn&&{} + 20.6694020171659560336324750416 \; \ep^{-2}
  \nn&&{} + 74.2503544994774685337366210644 \; \ep^{-1}
  \nn&&{} + 147.807870353419457006020027881
  \nn&&{} + 467.918626785158952385003318864 \; \ep
  \nn&&{} - 1988.53979380180044626102894892 \; \ep^{2}
  \nn&&{} - 4066.05004959291604399936393130 \; \ep^{3}
  \nn&&{} - 84210.5386674689601672799852884 \; \ep^{4}
  \\
  %
  % PR141[1, 1, 1, 1, 1, 1, 1, 0, 0, 0]
  \label{PR141}
  \masterpicture{
    %\begin{center}
    \fcolorbox{white}{white}{
      \SetScale{0.4}
      \SetOffset(25,0)
      %  \begin{picture}(147,150) (14,-11)
      \SetWidth{0.5}
      \SetColor{Black}
      \DashCArc(90,61)(74,90,217){4}                   % line 1
      \DashCArc(90,61)(74,-37,90){4}                   % line 2
      \CArc(90,61)(74,217,323)                  % line 3
      \DashCArc(120,95)(50,127,233){4}                 % line 4
      \DashCArc(60,95)(50,-53,53){4}                   % line 5
      \DashLine(90,55)(31,16.5){4}                     % line 6
      \DashLine(90,55)(149,16.5){4}                    % line 7
      \Vertex(90,135){2.83}
      \Vertex(90,55){2.83}
      \Vertex(149,16.5){2.83}
      \Vertex(31,16.5){2.83}
      %  \end{picture}
    }
    %\end{center}
  } &=&{} 
  + 0.0833333333333333333333333333333 \; \ep^{-4}
  \nn&&{} + 0.666666666666666666666666666667 \; \ep^{-3}
  \nn&&{} + 5.16908974465626417588448436109 \; \ep^{-2}
  \nn&&{} + 27.8538606812276657878535072006 \; \ep^{-1}
  \nn&&{} + 146.237985312171144034228066867
  \nn&&{} + 662.260137198260174836138241719 \; \ep
  \nn&&{} + 2974.21982572225816025721879182 \; \ep^{2}
  \nn&&{} + 12445.5794442004943383914555146 \; \ep^{3}
  \nn&&{} + 52217.5702319691751782305982084 \; \ep^{4}
  \\
  %
  % PR137[1, 1, 1, 1, 1, 1, 1, 0, 0, 0]
  \masterpicture{
    %\begin{center}
    \fcolorbox{white}{white}{
      \SetScale{0.4}
      \SetOffset(25,0)
      %  \begin{picture}(147,150) (14,-11)
      \SetWidth{0.5}
      \SetColor{Black}
      \DashCArc(90,61)(74,90,217){4}                   % line 1
      \DashCArc(90,61)(74,-37,90){4}                   % line 2
      \CArc(90,61)(74,217,323)                  % line 3
      \CArc(120,95)(50,127,233)                 % line 4
      \DashCArc(60,95)(50,-53,53){4}                   % line 5
      \Line(90,55)(31,16.5)                     % line 6
      \DashLine(90,55)(149,16.5){4}                    % line 7
      \Vertex(90,135){2.83}
      \Vertex(90,55){2.83}
      \Vertex(149,16.5){2.83}
      \Vertex(31,16.5){2.83}
      %  \end{picture}
    }
    %\end{center}
  } &=&{} 
  + 0.125000000000000000000000000000 \; \ep^{-4}
  \nn&&{} + 1.08333333333333333333333333333 \; \ep^{-3}
  \nn&&{} + 7.79306233504930030338751378907 \; \ep^{-2}
  \nn&&{} + 42.4229311299811602469885695910 \; \ep^{-1}
  \nn&&{} + 172.468639338411212673227289550
  \nn&&{} + 895.267223034507792336545934747 \; \ep
  \nn&&{} + 2851.24487745129920226056389169 \; \ep^{2}
  \nn&&{} + 15769.2924893036658768261052813 \; \ep^{3}
  \nn&&{} + 44094.4294409565473747922022723 \; \ep^{4}
  \\
  %
  % PR139[1, 1, 1, 1, 1, 1, 1, 0, 0, 0]
  \masterpicture{
    %\begin{center}
    \fcolorbox{white}{white}{
      \SetScale{0.4}
      \SetOffset(25,0)
      %  \begin{picture}(147,150) (14,-11)
      \SetWidth{0.5}
      \SetColor{Black}
      \DashCArc(90,61)(74,90,217){4}                   % line 1
      \CArc(90,61)(74,-37,90)                   % line 2
      \CArc(90,61)(74,217,323)                  % line 3
      \DashCArc(120,95)(50,127,233){4}                 % line 4
      \DashCArc(60,95)(50,-53,53){4}                   % line 5
      \Line(90,55)(31,16.5)                     % line 6
      \DashLine(90,55)(149,16.5){4}                    % line 7
      \Vertex(90,135){2.83}
      \Vertex(90,55){2.83}
      \Vertex(149,16.5){2.83}
      \Vertex(31,16.5){2.83}
      %  \end{picture}
    }
    %\end{center}
  } &=&{} 
  + 0.166666666666666666666666666667 \; \ep^{-4}
  \nn&&{} + 1.50000000000000000000000000000 \; \ep^{-3}
  \nn&&{} + 9.21497802228274214549080505555 \; \ep^{-2}
  \nn&&{} + 41.2601364660351459773071526496 \; \ep^{-1}
  \nn&&{} + 209.510736830949219399940244688
  \nn&&{} + 736.116832299866041378628875270 \; \ep
  \nn&&{} + 3742.45984331167422662482544777 \; \ep^{2}
  \nn&&{} + 11749.4302169340506083438761905 \; \ep^{3}
  \nn&&{} + 62286.5836001930375873157912315 \; \ep^{4}
  \\
  %
  % PR154[1, 1, 1, 1, 1, 1, 1, 0, 0, 0]
  \masterpicture{
    %\begin{center}
    \fcolorbox{white}{white}{
      \SetScale{0.4}
      \SetOffset(25,0)
      %  \begin{picture}(147,150) (14,-11)
      \SetWidth{0.5}
      \SetColor{Black}
      \DashCArc(90,61)(74,90,217){4}                   % line 1
      \CArc(90,61)(74,-37,90)                   % line 2
      \DashCArc(90,61)(74,217,323){4}                  % line 3
      \CArc(120,95)(50,127,233)                 % line 4
      \DashCArc(60,95)(50,-53,53){4}                   % line 5
      \Line(90,55)(31,16.5)                     % line 6
      \DashLine(90,55)(149,16.5){4}                    % line 7
      \Vertex(90,135){2.83}
      \Vertex(90,55){2.83}
      \Vertex(149,16.5){2.83}
      \Vertex(31,16.5){2.83}
      %  \end{picture}
    }
    %\end{center}
  } &=&{} 
  + 0.0833333333333333333333333333333 \; \ep^{-4}
  \nn&&{} + 0.833333333333333333333333333333 \; \ep^{-3}
  \nn&&{} + 7.93678486290272798525102010851 \; \ep^{-2}
  \nn&&{} + 36.6460666943862068010045914737 \; \ep^{-1}
  \nn&&{} + 194.097392655331631239559043545
  \nn&&{} + 770.178916823096887339876558212 \; \ep
  \nn&&{} + 3399.99482439741716780023304896 \; \ep^{2}
  \nn&&{} + 13282.1053505362611944372969993 \; \ep^{3}
  \nn&&{} + 54753.4641354130075585240962863 \; \ep^{4}
  \\
  %
  % PR154[1, 1, 1, 1, 1, 1, 2, 0, 0, 0]
  \masterpicture{
    %\begin{center}
    \fcolorbox{white}{white}{
      \SetScale{0.4}
      \SetOffset(25,0)
      %  \begin{picture}(147,150) (14,-11)
      \SetWidth{0.5}
      \SetColor{Black}
      \DashCArc(90,61)(74,90,217){4}                   % line 1
      \CArc(90,61)(74,-37,90)                   % line 2
      \DashCArc(90,61)(74,217,323){4}                  % line 3
      \CArc(120,95)(50,127,233)                 % line 4
      \DashCArc(60,95)(50,-53,53){4}                   % line 5
      \Line(90,55)(31,16.5)                     % line 6
      \DashLine(90,55)(149,16.5){4}                    % line 7
      \Vertex(90,135){2.83}
      \Vertex(90,55){2.83}
      \Vertex(149,16.5){2.83}
      \Vertex(31,16.5){2.83}
      \Vertex(119.5,35.75){5.66}
      %  \end{picture}
    }
    %\end{center}
  } &=&{} 
  - 0.125000000000000000000000000000 \; \ep^{-4}
  \nn&&{} - 0.750000000000000000000000000000 \; \ep^{-3}
  \nn&&{} - 2.71376648328794339088189620834 \; \ep^{-2}
  \nn&&{} - 7.09547330357733318621286446542 \; \ep^{-1}
  \nn&&{} - 22.2503497804982861883215445262
  \nn&&{} - 17.7874425940476975226672333232 \; \ep
  \nn&&{} - 175.317651447398940503032885592 \; \ep^{2}
  \nn&&{} + 313.983105353518264983361554717 \; \ep^{3}
  \nn&&{} - 2142.79466723532874375769322841 \; \ep^{4}
  \\
  %
  % PR147[1, 1, 1, 1, 1, 1, 1, 0, 0, 0]
  \masterpicture{
    %\begin{center}
    \fcolorbox{white}{white}{
      \SetScale{0.4}
      \SetOffset(25,0)
      %  \begin{picture}(147,150) (14,-11)
      \SetWidth{0.5}
      \SetColor{Black}
      \DashCArc(90,61)(74,90,217){4}                   % line 1
      \DashCArc(90,61)(74,-37,90){4}                   % line 2
      \DashCArc(90,61)(74,217,323){4}                  % line 3
      \CArc(120,95)(50,127,233)                 % line 4
      \CArc(60,95)(50,-53,53)                   % line 5
      \Line(90,55)(31,16.5)                     % line 6
      \Line(90,55)(149,16.5)                    % line 7
      \Vertex(90,135){2.83}
      \Vertex(90,55){2.83}
      \Vertex(149,16.5){2.83}
      \Vertex(31,16.5){2.83}
      %  \end{picture}
    }
    %\end{center}
  } &=&{} 
  + 0.0833333333333333333333333333333 \; \ep^{-4}
  \nn&&{} + 0.833333333333333333333333333333 \; \ep^{-3}
  \nn&&{} + 8.53781331448252512795088918927 \; \ep^{-2}
  \nn&&{} + 41.6650082807281404435406778114 \; \ep^{-1}
  \nn&&{} + 179.973949309303080544355066651
  \nn&&{} + 915.996742715562407821582232880 \; \ep
  \nn&&{} + 2767.39379170194861041563852821 \; \ep^{2}
  \nn&&{} + 16538.6792505904580566239289602 \; \ep^{3}
  \nn&&{} + 40456.0909713394095410013351009 \; \ep^{4}
  \\
  %
  % PR164[1, 1, 1, 1, 1, 1, 1, 0, 0, 0]
  \masterpicture{
    %\begin{center}
    \fcolorbox{white}{white}{
      \SetScale{0.4}
      \SetOffset(25,0)
      %  \begin{picture}(147,150) (14,-11)
      \SetWidth{0.5}
      \SetColor{Black}
      \DashCArc(90,61)(74,90,217){4}                   % line 1
      \CArc(90,61)(74,-37,90)                   % line 2
      \CArc(90,61)(74,217,323)                  % line 3
      \CArc(120,95)(50,127,233)                 % line 4
      \CArc(60,95)(50,-53,53)                   % line 5
      \DashLine(90,55)(31,16.5){4}                     % line 6
      \DashLine(90,55)(149,16.5){4}                    % line 7
      \Vertex(90,135){2.83}
      \Vertex(90,55){2.83}
      \Vertex(149,16.5){2.83}
      \Vertex(31,16.5){2.83}
      %  \end{picture}
    }
    %\end{center}
  } &=&{} 
  + 0.125000000000000000000000000000 \; \ep^{-4}
  \nn&&{} + 1.08333333333333333333333333333 \; \ep^{-3}
  \nn&&{} + 8.39409078662909744608738286983 \; \ep^{-2}
  \nn&&{} + 47.4418727163230938895246559287 \; \ep^{-1}
  \nn&&{} + 158.265894985614754765609093664
  \nn&&{} + 1040.51973958438238064321813619 \; \ep
  \nn&&{} + 2220.52649830986620272733311300 \; \ep^{2}
  \nn&&{} + 19009.3819367933781054485906380 \; \ep^{3}
  \nn&&{} + 29872.2890554813361688482086054 \; \ep^{4}
  \\
  %
  % PR174[1, 1, 1, 1, 1, 1, 1, 0, 0, 0]
  \masterpicture{
    %\begin{center}
    \fcolorbox{white}{white}{
      \SetScale{0.4}
      \SetOffset(25,0)
      %  \begin{picture}(147,150) (14,-11)
      \SetWidth{0.5}
      \SetColor{Black}
      \CArc(90,61)(74,90,217)                   % line 1
      \CArc(90,61)(74,-37,90)                   % line 2
      \CArc(90,61)(74,217,323)                  % line 3
      \CArc(120,95)(50,127,233)                 % line 4
      \DashCArc(60,95)(50,-53,53){4}                   % line 5
      \DashLine(90,55)(31,16.5){4}                     % line 6
      \DashLine(90,55)(149,16.5){4}                    % line 7
      \Vertex(90,135){2.83}
      \Vertex(90,55){2.83}
      \Vertex(149,16.5){2.83}
      \Vertex(31,16.5){2.83}
      %  \end{picture}
    }
    %\end{center}
  } &=&{} 
  + 0.166666666666666666666666666667 \; \ep^{-4}
  \nn&&{} + 1.50000000000000000000000000000 \; \ep^{-3}
  \nn&&{} + 9.81600647386253928819067413630 \; \ep^{-2}
  \nn&&{} + 46.2790780523770796198432389873 \; \ep^{-1}
  \nn&&{} + 192.143949985189792783133357259
  \nn&&{} + 895.583232707441960151372330064 \; \ep
  \nn&&{} + 3015.50609158685899428979534043 \; \ep^{2}
  \nn&&{} + 15431.7141565054206165629974866 \; \ep^{3}
  \nn&&{} + 45950.3851577110455999824698200 \; \ep^{4}
  \\
  %
  % PR193[1, 1, 1, 1, 1, 1, 1, 0, 0, 0]
  \masterpicture{
    %\begin{center}
    \fcolorbox{white}{white}{
      \SetScale{0.4}
      \SetOffset(25,0)
      %  \begin{picture}(147,150) (14,-11)
      \SetWidth{0.5}
      \SetColor{Black}
      \CArc(90,61)(74,90,217)                   % line 1
      \DashCArc(90,61)(74,-37,90){4}                   % line 2
      \DashCArc(90,61)(74,217,323){4}                  % line 3
      \CArc(120,95)(50,127,233)                 % line 4
      \DashCArc(60,95)(50,-53,53){4}                   % line 5
      \Line(90,55)(31,16.5)                     % line 6
      \Line(90,55)(149,16.5)                    % line 7
      \Vertex(90,135){2.83}
      \Vertex(90,55){2.83}
      \Vertex(149,16.5){2.83}
      \Vertex(31,16.5){2.83}
      %  \end{picture}
    }
    %\end{center}
  } &=&{} 
  + 0.125000000000000000000000000000 \; \ep^{-4}
  \nn&&{} + 1.25000000000000000000000000000 \; \ep^{-3}
  \nn&&{} + 9.95972900171596697005418045574 \; \ep^{-2}
  \nn&&{} + 41.2160203348045418858225510288 \; \ep^{-1}
  \nn&&{} + 210.851234758305984612090112605
  \nn&&{} + 790.910352868677999468381955470 \; \ep
  \nn&&{} + 3471.90742617992105936864664248 \; \ep^{2}
  \nn&&{} + 13388.2008481237206516463529138 \; \ep^{3}
  \nn&&{} + 54666.0539850126748720361226803 \; \ep^{4}
  \\
  %
  % PR158[1, 1, 1, 1, 1, 1, 1, 0, 0, 0]
  \masterpicture{
    %\begin{center}
    \fcolorbox{white}{white}{
      \SetScale{0.4}
      \SetOffset(25,0)
      %  \begin{picture}(147,150) (14,-11)
      \SetWidth{0.5}
      \SetColor{Black}
      \DashCArc(90,61)(74,90,217){4}                   % line 1
      \CArc(90,61)(74,-37,90)                   % line 2
      \DashCArc(90,61)(74,217,323){4}                  % line 3
      \CArc(120,95)(50,127,233)                 % line 4
      \CArc(60,95)(50,-53,53)                   % line 5
      \Line(90,55)(31,16.5)                     % line 6
      \Line(90,55)(149,16.5)                    % line 7
      \Vertex(90,135){2.83}
      \Vertex(90,55){2.83}
      \Vertex(149,16.5){2.83}
      \Vertex(31,16.5){2.83}
      %  \end{picture}
    }
    %\end{center}
  } &=&{} 
  + 0.125000000000000000000000000000 \; \ep^{-4}
  \nn&&{} + 1.25000000000000000000000000000 \; \ep^{-3}
  \nn&&{} + 10.5607574532957641127540495365 \; \ep^{-2}
  \nn&&{} + 46.2349619211464755283586373665 \; \ep^{-1}
  \nn&&{} + 188.248493841520722512481676436
  \nn&&{} + 965.746295254204695422785736581 \; \ep
  \nn&&{} + 2625.62618545640179663793065741 \; \ep^{2}
  \nn&&{} + 17532.0585981528153344707844000 \; \ep^{3}
  \nn&&{} + 36105.6708001198638661491047454 \; \ep^{4}
  \\
  %
  % PR145[1, 1, 1, 1, 1, 1, 1, 0, 0, 0]
  \masterpicture{
    %\begin{center}
    \fcolorbox{white}{white}{
      \SetScale{0.4}
      \SetOffset(25,0)
      %  \begin{picture}(147,150) (14,-11)
      \SetWidth{0.5}
      \SetColor{Black}
      \DashCArc(90,61)(74,90,217){4}                   % line 1
      \DashCArc(90,61)(74,-37,90){4}                   % line 2
      \CArc(90,61)(74,217,323)                  % line 3
      \CArc(120,95)(50,127,233)                 % line 4
      \CArc(60,95)(50,-53,53)                   % line 5
      \DashLine(90,55)(31,16.5){4}                     % line 6
      \DashLine(90,55)(149,16.5){4}                    % line 7
      \Vertex(90,135){2.83}
      \Vertex(90,55){2.83}
      \Vertex(149,16.5){2.83}
      \Vertex(31,16.5){2.83}
      %  \end{picture}
    }
    %\end{center}
  } &=&{} 
  + 0.0833333333333333333333333333333 \; \ep^{-4}
  \nn&&{} + 0.666666666666666666666666666667 \; \ep^{-3}
  \nn&&{} + 6.37114664781585928445076208433 \; \ep^{-2}
  \nn&&{} + 37.8917438539115415433555091730 \; \ep^{-1}
  \nn&&{} + 154.604631544617018499470855434
  \nn&&{} + 894.807380418881266203622406485 \; \ep
  \nn&&{} + 2607.13652650580177242514674361 \; \ep^{2}
  \\
  %
  % PR145[1, 1, 1, 1, 1, 1, 2, 0, 0, 0]
  \masterpicture{
    %\begin{center}
    \fcolorbox{white}{white}{
      \SetScale{0.4}
      \SetOffset(25,0)
      %  \begin{picture}(147,150) (14,-11)
      \SetWidth{0.5}
      \SetColor{Black}
      \DashCArc(90,61)(74,90,217){4}                   % line 1
      \DashCArc(90,61)(74,-37,90){4}                   % line 2
      \CArc(90,61)(74,217,323)                  % line 3
      \CArc(120,95)(50,127,233)                 % line 4
      \CArc(60,95)(50,-53,53)                   % line 5
      \DashLine(90,55)(31,16.5){4}                     % line 6
      \DashLine(90,55)(149,16.5){4}                    % line 7
      \Vertex(90,135){2.83}
      \Vertex(90,55){2.83}
      \Vertex(149,16.5){2.83}
      \Vertex(31,16.5){2.83}
      \Vertex(90,-13){5.66}
      %  \end{picture}
    }
    %\end{center}
  } &=&{} 
  + 0.0833333333333333333333333333333 \; \ep^{-4}
  \nn&&{} + 0.333333333333333333333333333333 \; \ep^{-3}
  \nn&&{} + 2.50242307798959750921781769442 \; \ep^{-2}
  \nn&&{} + 7.17750170260260867273229997543 \; \ep^{-1}
  \nn&&{} + 10.0033662615671539152667896043
  \nn&&{} + 121.389726250269120381127113931 \; \ep
  \nn&&{} - 338.169587547191526007167931728 \; \ep^{2}
  \\
  %
  % PR177[1, 1, 1, 1, 1, 1, 1, 1, 0, 0]
  \masterpicture{
    %\begin{center}
    \fcolorbox{white}{white}{
      \SetScale{0.3}
      \SetOffset(-5,-5)
      %  \begin{picture}(310,276) (0,0)
      \SetWidth{0.5}
      \SetColor{Black}
      \DashCArc(218,100)(38,90,270){4}
      \DashCArc(218,100)(38,-90,90){4}
      \DashCArc(218,100)(98,90,235){4}
      \CArc(218,100)(98,235,305)
      \CArc(218,100)(98,-55,90)
      \Line(218,198)(218,138)
      \DashLine(218,62)(161.8,19.7){4}
      \DashLine(218,62)(274.2,19.7){4}
      \Vertex(218,198){2.83}
      \Vertex(218,138){2.83}
      \Vertex(218,62){2.83}
      \Vertex(161.8,19.7){2.83}
      \Vertex(274.2,19.7){2.83}
      %  \end{picture}
    }
    %\end{center}
  } &=&{} 
  + 1.80308535473939142809960724227 \; \ep^{-2}
  \nn&&{} - 2.03566656453161423532377023190 \; \ep^{-1}
  \nn&&{} + 40.4773901150829247806609147362
  \nn&&{} - 100.628977840966481693832556086 \; \ep
  \nn&&{} + 679.056339067891403282049708894 \; \ep^{2}
  \nn&&{} - 2387.60452082758635632061839710 \; \ep^{3}
  \\
  %
  % PR179[1, 1, 1, 1, 1, 1, 1, 1, 0, 0]
  \masterpicture{
    %\begin{center}
    \fcolorbox{white}{white}{
      \SetScale{0.48}
      \SetOffset(21,-3)
      %  \begin{picture}(154,149) (0,0)
      \SetWidth{0.5}
      \SetColor{Black}
      \DashCArc(85,66)(62,90,180){4}          % line 1
      \CArc(85,66)(62,0,90)            % line 2
      \CArc(85,66)(62,-90,0)           % line 3
      \DashCArc(85,66)(62,180,270){4}         % line 4
      \DashLine(23,66)(85,66){4}              % line 5
      \DashLine(85,128)(85,66){4}             % line 6
      \DashLine(147,66)(85,66){4}             % line 7
      \DashLine(85,4)(85,66){4}               % line 8
      \Vertex(85,66){2.83}
      \Vertex(85,128){2.83}
      \Vertex(23,66){2.83}
      \Vertex(85,4){2.83}
      \Vertex(147,66){2.83}
      %  \end{picture}
    }
    %\end{center}
  } &=&{} 
  + 5.18463877571684963165682743229 \; \ep^{-1}
  \nn&&{} + 0.374844191926851274910805165871
  \nn&&{} + 141.683133328263640451220557622 \; \ep
  \nn&&{} - 146.684014785607112711629455843 \; \ep^{2}
  \nn&&{} + 2448.77867872444787260224256255 \; \ep^{3}
  \\
  %
  % PR180[1, 1, 1, 1, 1, 1, 1, 1, 0, 0]
  \masterpicture{
    %\begin{center}
    \fcolorbox{white}{white}{
      \SetScale{0.48}
      \SetOffset(21,-3)
      %  \begin{picture}(154,149) (0,0)
      \SetWidth{0.5}
      \SetColor{Black}
      \CArc(85,66)(62,90,180)          % line 1
      \DashCArc(85,66)(62,0,90){4}            % line 2
      \CArc(85,66)(62,-90,0)           % line 3
      \CArc(85,66)(62,180,270)         % line 4
      \DashLine(23,66)(85,66){4}              % line 5
      \DashLine(85,128)(85,66){4}             % line 6
      \DashLine(147,66)(85,66){4}             % line 7
      \DashLine(85,4)(85,66){4}               % line 8
      \Vertex(85,66){2.83}
      \Vertex(85,128){2.83}
      \Vertex(23,66){2.83}
      \Vertex(85,4){2.83}
      \Vertex(147,66){2.83}
      %  \end{picture}
    }
    %\end{center}
  } &=&{} 
  + 5.18463877571684963165682743229 \; \ep^{-1}
  \nn&&{} - 13.7206302625920162070487342659
  \nn&&{} + 135.928906108952872038638377087 \; \ep
  \nn&&{} - 497.764644552772376477970102393 \; \ep^{2}
  \nn&&{} + 2695.35125395194610632909729485 \; \ep^{3}
  \\
  %
  % PR186[1, 1, 1, 1, 1, 1, 1, 1, 0, 0]
  \masterpicture{
    %\begin{center}
    \fcolorbox{white}{white}{
      \SetScale{0.48}
      \SetOffset(21,-3)
      %  \begin{picture}(154,149) (0,0)
      \SetWidth{0.5}
      \SetColor{Black}
      \DashCArc(85,66)(62,90,180){4}          % line 1
      \DashCArc(85,66)(62,0,90){4}            % line 2
      \CArc(85,66)(62,-90,0)           % line 3
      \DashCArc(85,66)(62,180,270){4}         % line 4
      \DashLine(23,66)(85,66){4}              % line 5
      \Line(85,128)(85,66)             % line 6
      \Line(147,66)(85,66)             % line 7
      \DashLine(85,4)(85,66){4}               % line 8
      \Vertex(85,66){2.83}
      \Vertex(85,128){2.83}
      \Vertex(23,66){2.83}
      \Vertex(85,4){2.83}
      \Vertex(147,66){2.83}
      %  \end{picture}
    }
    %\end{center}
  } &=&{} 
  + 5.18463877571684963165682743229 \; \ep^{-1}
  \nn&&{} - 7.85876279518922242360822990783
  \nn&&{} + 128.531386676431303444223252317 \; \ep
  \nn&&{} - 349.879176215675436895102723351 \; \ep^{2}
  \nn&&{} + 2332.83399433790756728143973285 \; \ep^{3}
  \\
  %
  % PR196[1, 1, 1, 1, 1, 1, 1, 1, 0, 0]
  \masterpicture{
    %\begin{center}
    \fcolorbox{white}{white}{
      \SetScale{0.48}
      \SetOffset(21,-3)
      %  \begin{picture}(154,149) (0,0)
      \SetWidth{0.5}
      \SetColor{Black}
      \DashCArc(85,66)(62,90,180){4}          % line 1
      \CArc(85,66)(62,0,90)            % line 2
      \CArc(85,66)(62,-90,0)           % line 3
      \CArc(85,66)(62,180,270)         % line 4
      \DashLine(23,66)(85,66){4}              % line 5
      \Line(85,128)(85,66)             % line 6
      \DashLine(147,66)(85,66){4}             % line 7
      \DashLine(85,4)(85,66){4}               % line 8
      \Vertex(85,66){2.83}
      \Vertex(85,128){2.83}
      \Vertex(23,66){2.83}
      \Vertex(85,4){2.83}
      \Vertex(147,66){2.83}
      %  \end{picture}
    }
    %\end{center}
  } &=&{} 
  + 5.18463877571684963165682743229 \; \ep^{-1}
  \nn&&{} - 20.8585976739540823192137169924
  \nn&&{} + 148.814327289184790880960752284 \; \ep
  \nn&&{} - 657.418920712168196526528768117 \; \ep^{2}
  \nn&&{} + 3182.47221782449086620274231596 \; \ep^{3}
  \\
  %
  % PR201[1, 1, 1, 1, 1, 1, 1, 1, 0, 0]
  \masterpicture{
    %\begin{center}
    \fcolorbox{white}{white}{
      \SetScale{0.48}
      \SetOffset(21,-3)
      %  \begin{picture}(154,149) (0,0)
      \SetWidth{0.5}
      \SetColor{Black}
      \DashCArc(85,66)(62,90,180){4}          % line 1
      \CArc(85,66)(62,0,90)            % line 2
      \DashCArc(85,66)(62,-90,0){4}           % line 3
      \CArc(85,66)(62,180,270)         % line 4
      \DashLine(23,66)(85,66){4}              % line 5
      \Line(85,128)(85,66)             % line 6
      \Line(147,66)(85,66)             % line 7
      \DashLine(85,4)(85,66){4}               % line 8
      \Vertex(85,66){2.83}
      \Vertex(85,128){2.83}
      \Vertex(23,66){2.83}
      \Vertex(85,4){2.83}
      \Vertex(147,66){2.83}
      %  \end{picture}
    }
    %\end{center}
  } &=&{} 
  + 5.18463877571684963165682743229 \; \ep^{-1}
  \nn&&{} - 18.7699726985410818615637417243
  \nn&&{} + 142.191941064595949485071666672 \; \ep
  \nn&&{} - 600.255511127362803430335175869 \; \ep^{2}
  \nn&&{} + 2957.18558813594868083060620764 \; \ep^{3}
  \\
  %
  % PR203[1, 1, 1, 1, 1, 1, 1, 1, 0, 0]
  \masterpicture{
    %\begin{center}
    \fcolorbox{white}{white}{
      \SetScale{0.48}
      \SetOffset(21,-3)
      %  \begin{picture}(154,149) (0,0)
      \SetWidth{0.5}
      \SetColor{Black}
      \CArc(85,66)(62,90,180)          % line 1
      \DashCArc(85,66)(62,0,90){4}            % line 2
      \CArc(85,66)(62,-90,0)           % line 3
      \DashCArc(85,66)(62,180,270){4}         % line 4
      \Line(23,66)(85,66)              % line 5
      \DashLine(85,128)(85,66){4}             % line 6
      \Line(147,66)(85,66)             % line 7
      \DashLine(85,4)(85,66){4}               % line 8
      \Vertex(85,66){2.83}
      \Vertex(85,128){2.83}
      \Vertex(23,66){2.83}
      \Vertex(85,4){2.83}
      \Vertex(147,66){2.83}
      %  \end{picture}
    }
    %\end{center}
  } &=&{} 
  + 5.18463877571684963165682743229 \; \ep^{-1}
  \nn&&{} - 19.0793750927079960314257405876
  \nn&&{} + 141.252248186107747164092632797 \; \ep
  \nn&&{} - 605.029621201870258904662671455 \; \ep^{2}
  \nn&&{} + 2946.48740435117339409606564510 \; \ep^{3}
  \\
  %
  % PR208[1, 1, 1, 1, 1, 1, 1, 1, 0, 0]
  \masterpicture{
    %\begin{center}
    \fcolorbox{white}{white}{
      \SetScale{0.48}
      \SetOffset(21,-3)
      %  \begin{picture}(154,149) (0,0)
      \SetWidth{0.5}
      \SetColor{Black}
      \DashCArc(85,66)(62,90,180){4}          % line 1
      \CArc(85,66)(62,0,90)            % line 2
      \CArc(85,66)(62,-90,0)           % line 3
      \DashCArc(85,66)(62,180,270){4}         % line 4
      \Line(23,66)(85,66)              % line 5
      \DashLine(85,128)(85,66){4}             % line 6
      \DashLine(147,66)(85,66){4}             % line 7
      \Line(85,4)(85,66)               % line 8
      \Vertex(85,66){2.83}
      \Vertex(85,128){2.83}
      \Vertex(23,66){2.83}
      \Vertex(85,4){2.83}
      \Vertex(147,66){2.83}
      %  \end{picture}
    }
    %\end{center}
  } &=&{} 
  + 5.18463877571684963165682743229 \; \ep^{-1}
  \nn&&{} - 19.1118657059490732311103013898
  \nn&&{} + 141.144175505094858012998872298 \; \ep
  \nn&&{} - 605.431190575217055408942299967 \; \ep^{2}
  \nn&&{} + 2945.58041293149149762720437022 \; \ep^{3}
  \\
  %
  % PR191[1, 1, 1, 1, 1, 1, 1, 1, 0, 0]
  \masterpicture{
    %\begin{center}
    \fcolorbox{white}{white}{
      \SetScale{0.48}
      \SetOffset(21,-3)
      %  \begin{picture}(154,149) (0,0)
      \SetWidth{0.5}
      \SetColor{Black}
      \DashCArc(85,66)(62,90,180){4}          % line 1
      \DashCArc(85,66)(62,0,90){4}            % line 2
      \DashCArc(85,66)(62,-90,0){4}           % line 3
      \DashCArc(85,66)(62,180,270){4}         % line 4
      \DashLine(23,66)(85,66){4}              % line 5
      \Line(85,128)(85,66)             % line 6
      \DashLine(147,66)(85,66){4}             % line 7
      \Line(85,4)(85,66)               % line 8
      \Vertex(85,66){2.83}
      \Vertex(85,128){2.83}
      \Vertex(23,66){2.83}
      \Vertex(85,4){2.83}
      \Vertex(147,66){2.83}
      %  \end{picture}
    }
    %\end{center}
  } &=&{} 
  + 5.18463877571684963165682743229 \; \ep^{-1}
  \nn&&{} + 7.30445968508659934705130294982
  \nn&&{} + 141.443417856878944115040221346 \; \ep
  \nn&&{} + 40.4773793831579631350040651322 \; \ep^{2}
  \\
  %
  % PR215[1, 1, 1, 1, 1, 1, 1, 1, 1, 0]
  \masterpicture{
    %\begin{center}
    \fcolorbox{white}{white}{
      \SetScale{0.35}
      \SetOffset(-50,25)
      %  \begin{picture}(178,156) (219,-69)
      \SetWidth{0.5}
      \SetColor{Black}
      \DashLine(265,82)(353,82){4}           % line 1
      \DashLine(353,82)(395,8){4}            % line 2
      \DashLine(353,-66)(395,8){4}           % line 3
      \Line(265,-66)(353,-66)         % line 4
      \Line(223,8)(265,-66)           % line 5
      \Line(265,82)(223,8)            % line 6
      \DashLine(265,82)(304,16){4}           % line 7a
      \DashLine(314,4)(353,-66){4}           % line 7b
      \DashLine(353,82)(265,-66){4}          % line 8
      \DashLine(223,8)(296,8){4}             % line 9a
      \DashLine(330,8)(395,8){4}             % line 9b
      \Vertex(265,82){2.83}
      \Vertex(223,8){2.83}
      \Vertex(265,-66){2.83}
      \Vertex(353,-66){2.83}
      \Vertex(353,82){2.83}
      \Vertex(395,8){2.83}
      %  \end{picture}
    }
    %\end{center}
  } &=&{} 
  - 6.72847056008568105547188977521
  \nn&&{} - 26.0876465999666155389659770717 \; \ep
  \nn&&{} - 214.647717912411362028052727052 \; \ep^{2}
  \nn&&{} - 613.715203096626075654874908838 \; \ep^{3}
  \\
  %
  % PR202[1, 1, 1, 1, 1, 1, 1, 1, 1, 0]
  \masterpicture{
    %\begin{center}
    \fcolorbox{white}{white}{
      \SetScale{0.35}
      \SetOffset(-50,25)
      %  \begin{picture}(178,156) (219,-69)
      \SetWidth{0.5}
      \SetColor{Black}
      \DashLine(265,82)(353,82){4}           % line 1
      \Line(353,82)(395,8)            % line 2
      \Line(353,-66)(395,8)           % line 3
      \Line(265,-66)(353,-66)         % line 4
      \Line(223,8)(265,-66)           % line 5
      \DashLine(265,82)(223,8){4}            % line 6
      \DashLine(265,82)(304,16){4}           % line 7a
      \DashLine(314,4)(353,-66){4}           % line 7b
      \DashLine(353,82)(265,-66){4}          % line 8
      \DashLine(223,8)(296,8){4}             % line 9a
      \DashLine(330,8)(395,8){4}             % line 9b
      \Vertex(265,82){2.83}
      \Vertex(223,8){2.83}
      \Vertex(265,-66){2.83}
      \Vertex(353,-66){2.83}
      \Vertex(353,82){2.83}
      \Vertex(395,8){2.83}
      %  \end{picture}
    }
    %\end{center}
  } &=&{} 
  - 3.71140264536682392682628965373
  \nn&&{} - 2.11520599545877264756135261804 \; \ep
  \nn&&{} - 71.9899451389829491830674629550 \; \ep^{2}
  \nn&&{} + 41.1881294174309244417864419468 \; \ep^{3}
  \\
  %
  % PR224[1, 1, 1, 1, 1, 1, 1, 1, 1, 0]
  \label{PR224}
  \masterpicture{
    %\begin{center}
    \fcolorbox{white}{white}{
      \SetScale{0.35}
      \SetOffset(-50,25)
      %  \begin{picture}(178,156) (219,-69)
      \SetWidth{0.5}
      \SetColor{Black}
      \Line(265,82)(353,82)                  % line 1
      \DashLine(353,82)(395,8){4}            % line 2
      \DashLine(353,-66)(395,8){4}           % line 3
      \DashLine(265,-66)(353,-66){4}         % line 4
      \DashLine(223,8)(265,-66){4}           % line 5
      \DashLine(265,82)(223,8){4}            % line 6
      \DashLine(265,82)(304,16){4}           % line 7a
      \DashLine(314,4)(353,-66){4}           % line 7b
      \DashLine(353,82)(265,-66){4}          % line 8
      \DashLine(223,8)(296,8){4}             % line 9a
      \DashLine(330,8)(395,8){4}             % line 9b
      \Vertex(265,82){2.83}
      \Vertex(223,8){2.83}
      \Vertex(265,-66){2.83}
      \Vertex(353,-66){2.83}
      \Vertex(353,82){2.83}
      \Vertex(395,8){2.83}
      %  \end{picture}
    }
    %\end{center}
  } &=&{} 
  + 5.18463877571684963165682743229  \; \ep^{-1}
  \nn&&{} + 52.8346981753279451179346027590
  \nn&&{} + 447.323275771878352225930613955 \; \ep
  \\
  %
  % PR221[1, 1, 1, 1, 1, 1, 1, 1, 1, 0]
  \masterpicture{
    %\begin{center}
    \fcolorbox{white}{white}{
      \SetScale{0.42}
      \SetOffset(-5,35)
      %  \begin{picture}(158,111) (73,-71)
      \SetWidth{0.5}
      \SetColor{Black}
      \Line(150,35)(76,-69)
      \Line(150,37)(229,-69)
      \Line(150,-9)(120,-53)
      \Line(150,-9)(180,-53)
      \Line(180,-53)(230,-69)
      \DashLine(77,-69)(230,-69){4}
      \DashLine(150,38)(150,-9){4}
      \DashLine(120,-53)(77,-69){4}
      \DashLine(120,-53)(180,-53){4}
      \Vertex(77,-69){2.83}
      \Vertex(229,-69){2.83}
      \Vertex(150,36){2.83}
      \Vertex(120,-53){2.83}
      \Vertex(180,-53){2.83}
      \Vertex(150,-9){2.83}
      %  \end{picture}
    }
    %\end{center}
  } &=&{} 
  - 3.4497511317390349922288
  \nn&&{} + 6.3127694885459812824115 \; \ep
  \nn&&{} - 63.668771344187502234181 \; \ep^{2}
  \nn&&{} + 196.34402612627359322923 \; \ep^{3}
  \\
  %
  % PR219[1, 1, 1, 1, 1, 1, 1, 1, 1, 0]
  \masterpicture{
    %\begin{center}
    \fcolorbox{white}{white}{
      \SetScale{0.35}
      \SetOffset(-50,25)
      %  \begin{picture}(178,156) (219,-69)
      \SetWidth{0.5}
      \SetColor{Black}
      \DashLine(265,82)(353,82){4}           % line 1
      \Line(353,82)(395,8)            % line 2
      \Line(353,-66)(395,8)           % line 3
      \Line(265,-66)(353,-66)         % line 4
      \Line(223,8)(265,-66)           % line 5
      \Line(265,82)(223,8)            % line 6
      \DashLine(265,82)(304,16){4}           % line 7a
      \DashLine(314,4)(353,-66){4}           % line 7b
      \DashLine(353,82)(265,-66){4}          % line 8
      \DashLine(223,8)(296,8){4}             % line 9a
      \DashLine(330,8)(395,8){4}             % line 9b
      \Vertex(265,82){2.83}
      \Vertex(223,8){2.83}
      \Vertex(265,-66){2.83}
      \Vertex(353,-66){2.83}
      \Vertex(353,82){2.83}
      \Vertex(395,8){2.83}
      %  \end{picture}
    }
    %\end{center}
  } &=&{} 
  - 2.42695639537700735
  \nn&&{} + 4.01554669961524192 \; \ep
  \nn&&{} - 43.6962533603324647 \; \ep^{2}
  \nn&&{} + 128.936157875347890 \; \ep^{3}
  \\
  %
  % PR219[1, 1, 1, 1, 1, 1, 1, 1, 2, 0]
  \masterpicture{
    %\begin{center}
    \fcolorbox{white}{white}{
      \SetScale{0.35}
      \SetOffset(-50,25)
      %  \begin{picture}(178,156) (219,-69)
      \SetWidth{0.5}
      \SetColor{Black}
      \DashLine(265,82)(353,82){4}           % line 1
      \Line(353,82)(395,8)            % line 2
      \Line(353,-66)(395,8)           % line 3
      \Line(265,-66)(353,-66)         % line 4
      \Line(223,8)(265,-66)           % line 5
      \Line(265,82)(223,8)            % line 6
      \DashLine(265,82)(304,16){4}           % line 7a
      \DashLine(314,4)(353,-66){4}           % line 7b
      \DashLine(353,82)(265,-66){4}          % line 8
      \DashLine(223,8)(296,8){4}             % line 9a
      \DashLine(330,8)(395,8){4}             % line 9b
      \Vertex(265,82){2.83}
      \Vertex(223,8){2.83}
      \Vertex(265,-66){2.83}
      \Vertex(353,-66){2.83}
      \Vertex(353,82){2.83}
      \Vertex(395,8){2.83}
      \Vertex(244,-29){5.66}
      %  \end{picture}
    }
    %\end{center}
  } &=&{} 
  + 0.473611472272364450
  \nn&&{} + 1.09585342206826990 \; \ep
  \nn&&{} + 5.37764333252884269 \; \ep^{2}
  \nn&&{} + 8.82896457590640998 \; \ep^{3}
  %\\
  \eea

  Let us finally note that the prototypes in Eqs.~\ref{PR141} and \ref{PR224}
  contain only one massive line, and are therefore trivially amenable
  to massless three-loop propagators. In order to evaluate these, we
  have used the results from
  \cite{Bierenbaum:2003ud,Kazakov:1983ns,Bekavac:2005xs}. For example
  the result in Eq.~\ref{PR224} can be obtained from

  \bea
  &&\!\!\!\!\!\!-e^{4\ep \gamma}\frac{\Gamma(-4\ep)\Gamma^3(1 + \ep)\Gamma^6(1 -
    \ep)\Gamma(1 + 4\ep)}{\Gamma^3(2 - 2\ep)\Gamma(2 - \ep)}
  \left(20\zeta_5 + (68\zeta_3^2 - 80\zeta_5+50\zeta_6)\;\ep \right. \nn &&+
  \left. (-272\zeta_3^2 + 204 \zeta_3\zeta_4 + 80\zeta_5 - 200 \zeta_6 +
  450\zeta_7)\;\ep^2 + O(\ep^3)\right)
  \eea

  Although not all of the above numerical values could be checked with
  independent techniques, we were able to verify many of them, either
  with the sector decomposition method \cite{sectors}, or with
  Mellin-Barnes techniques \cite{Smirnov:1999gc,Tausk:1999vh}
  implemented in the {\tt MB} package \cite{Czakon:2005rk} (see also
  \cite{Anastasiou:2005cb}).

  \section{On-shell Results}

  \label{results}

  Once the master integrals are calculated, it is just a matter of
  substituting their values into the reduced diagrams and performing
  $\msb$ renormalization to obtain the corrections to
  $\Delta \rho$ in the $\msb$ scheme. The latter can be found in the Appendix. As far as
  applications to electroweak physics are concerned, the on-shell
  scheme is more relevant. We used the relation between the $\msb$ and
  on-shell masses of a heavy quark from \cite{Melnikov:2000qh}, in
  order to perform the translation between the schemes.
  
  In our result below, we keep the explicit dependence on the
  logarithms of the ratio of the on-shell top quark mass and the dimensional
  regularization mass unit. These can be checked, or recovered if not
  given, by using the simple renormalization group equation satisfied
  by $\delta \rho$

  \be
  0 = \frac{d}{d \log(\mu^2)} \; \delta \rho \; ,
  \ee
  
  which in the on-shell scheme does only require the knowledge of the
  QCD $\beta$-function.Clearly, in the $\msb$ scheme the anomalous
  dimension of the mass would also be needed.

  If we introduce the following notation
 
  \be
  X_t = \frac{G_F m_t^2}{8\sqrt{2}\pi^2}, \;\;\;\;
  L_t \equiv \log\bigg(\frac{m_t^2}{\mu^2}\bigg),
  \ee

  the QCD corrections up to ${\mathcal O}(\as^3)$ to the leading-order $\Delta
  \rho$ result in the on-shell scheme are

  \bea
  &&\delta \rho^{OS}(\mbox{non-singlet}) = 3 X_t \,
  \Bigg\{ 
  1 +  \bigg( \frac{\as}{\pi}\bigg) \,C_F\, 
  \bigg( - 2.144934067 \bigg)
  \\ &&
  \hspace{0.25cm}  +  \bigg( \frac{\as}{\pi}\bigg)^2 \,
  \Bigg[\hspace{.1cm} C_F^2 \,\bigg( 3.228455773 \bigg)
    \nonumber \\ &&
    \hspace{0.5cm} {}  
    +  C_F\,C_A \,\bigg(-6.288851333 
    + 1.966189561 \,L_t \bigg) 
    \nonumber \\ &&
    \hspace{0.5cm} {}  
    +  C_F\,T_F\,\bigg(0.1470525995 
    - 0.7149780223 \,L_t 
    \nonumber \\ &&
    \hspace{1.0cm} {} 
    + n_l\,\bigg(2.679319209 
    - 0.7149780223 \, L_t \bigg)
    \bigg)  \Bigg]
  \nonumber \\ &&
  \hspace{0.25cm}  +  \bigg( \frac{\as}{\pi}\bigg)^3 \, 
  \Bigg[\hspace{.1cm}\, C_F^3\,\bigg( -0.7845479837 \bigg)
    \nonumber \\ &&
    \hspace{0.5cm} {}  
    + C_F^2\,C_A\,\bigg(17.20096563  
    - 5.918835584 \, L_t \bigg) 
    \nonumber \\ &&
    \hspace{0.5cm} {}   + C_F^2\,T_F\,\bigg(-0.4393186129 
    + 1.616070332 \,L_t 
    \nonumber \\ &&
    \hspace{1.0cm} {} 
    + n_l\,\bigg(-8.740003239 
    + 1.616070332 \, L_t \bigg) \bigg)
    \nonumber \\ &&
    \hspace{0.5cm} {}
    + C_F\,C_A^2\,\bigg(-30.95679757  
    + 13.0488891 \,L_t 
    - 1.802340431 \,L_t^2 \bigg)
    \nonumber \\ &&
    \hspace{0.5cm} {} 
    + C_F\,C_A\,T_F\,\bigg(-0.5400590182 
    - 5.355886515 \, L_t 
    + 1.310793041\, L_t^2 
    \nonumber \\ &&
    \hspace{1.0cm} {} 
    + n_l\,\bigg(24.8274162 
    - 9.998375300 \,L_t 
    + 1.310793041 \,L_t^2 \bigg) \bigg)
    \nonumber \\ &&
    \hspace{0.5cm} {} 
    + C_F\,T_F^2 \, \bigg(0.3035659457 
    + 0.09803506636 \,L_t 
    - 0.2383260074 \, L_t^2 
    \nonumber \\ &&
    \hspace{1.0cm} {}   
    + n_l\,\bigg(0.7160711769 
    + 1.884247873 \,L_t 
    - 0.4766520149 \,L_t^2 \bigg) 
    \nonumber \\ &&
    \hspace{1.0cm} {}   
    + n_l^2\,\bigg(-3.448039206  
    +1.786212806 \,L_t 
    -0.2383260074 \, L_t^2 \bigg) \bigg)
    \Bigg]
  \Bigg\}
  \nonumber
  \eea

  After setting $\mu = m_t$, which is equivalent to $L_t = 0$, our
  result is in agreement with \cite{Chetyrkin:2006bj}.

  \section{Conclusions}

  In this work, we have computed a further subset of four-loop single scale
  vacuum integrals with enough terms in the $\ep$ expansion to even
  allow for five-loop calculations. Our results can be applied
  whenever the physical process shows large scale differences and the
  large mass procedure can be applied, leading naturally to expansion
  (Wilson) coefficients expressed through tadpoles.

  The immediate motivation for this computation has been the
  calculation of the non-singlet corrections to the $\rho$ parameter
  at the four-loop level. After completing this task, we have obtained
  a result in agreement with the one recently presented in
  \cite{Chetyrkin:2006bj}. Thus, the complete four-loop corrections
  are now available and checked by two independent calculations for
  both the singlet (\cite{Schroder:2005db} and
  \cite{Chetyrkin:2006bj}) and non-singlet
  (\cite{Chetyrkin:2006bj} and the present work) parts. We can, of
  course, only confirm the numerical smallness of the final
  correction.
  
  \section{Acknowledgments}
  
  Parts of the presented calculations were performed on the DESY
  Zeuthen Grid Engine computer cluster. This work was supported by the
  Sofja Kovalevskaja Award of the Alexander von Humboldt Foundation
  sponsored by the German Federal Ministry of Education and Research.
  
  \appendix

  \section{The $\rho$ Parameter in the $\msb$ Scheme}

  The substitution of the results of Section~\ref{masters} into the
  expression for the sum of the four-loop diagrams after reduction to masters
  leads to the bare contribution. Together with lower order
  corrections it can be written as follows

  \be
  \delta \rho = \frac{G_F m_0^2}{8\sqrt{2}\pi^2}
  \left( \frac{m_0}{\mu} \right)^{-2\ep} \sum_{i=0}^\infty
  \left( \frac{\as^0}{\pi} m_0^{-2\ep} \right)^i c_i(\ep),
  \ee

  where $c_i(\ep)$ are given as Laurent expansions in $\ep$, with expansion
  coefficients being pure numbers. We perform the $\msb$ renormalization with

  \be
  \as^0 = \mu^{2\ep} Z_{\as} \as(\mu^2),
  \ee
  \be
  m_0 = Z_m \mtbar(\mu^2),
  \ee

  where we only need the two-loop strong coupling
  renormalization constant and the three-loop mass renormalization
  constant. With the following definitions

  \be
  x_t(\mu^2) = \frac{G_F \mtbar^2(\mu^2)}{8\sqrt{2}\pi^2}, \;\;\;\;
  l_t \equiv \log\bigg(\frac{\mtbar^2(\mu^2)}{\mu^2}\bigg),
  \ee

  the contribution to $\Delta \rho$ of the QCD dressed one-loop
  diagrams up to ${\mathcal O}(\as^3)$ can be written as\footnote{dressing the
  one-loop diagrams cannot generate singlet contributions.}

  \bea
  \label{eq:MSbar-rho}
  && \delta \rho^{\overline{\mathrm{MS}}}(\mbox{non-singlet}) = 3 x_t \,
  \Bigg\{ 
  1 +  \bigg( \frac{\as}{\pi}\bigg) \,C_F\, 
  \bigg(\, -0.1449340668 - 1.5 \,\, l_t  \,\bigg)
  \\ &&
  \hspace{0.25cm} +   \bigg( \frac{\as}{\pi}\bigg)^2 \,
  \Bigg[\hspace{.1cm} C_F^2 \,\bigg(-0.04028771897 
    + 2.279901100 \, l_t 
    + 1.125 \, l_t^2 \bigg) 
    \nonumber \\ &&
    \hspace{0.5cm} {}             + C_F \,C_A \,\bigg( 0.3716744884 
    - 1.887977105 \, l_t 
    + 0.6875 \, l_t^2 \bigg)
    \nonumber \\ && 
    \hspace{0.5cm} {}             + C_F \,T_F\,\bigg(0.4577540666 
    + 0.3683553111 \, l_t 
    - 0.25 \, l_t^2
    \nonumber \\ && 
    \hspace{1.0cm} {}             + n_l\,\bigg(-0.4447815241 
    + 0.3683553111 \, l_t 
    -0.25 \, l_t^2 \bigg) \bigg) 
    \Bigg] 
  \nonumber \\ &&
  \hspace{0.25cm} +   \bigg( \frac{\as}{\pi}\bigg)^3 \,
  \Bigg[ \hspace{.1cm}  C_F^3 \,\bigg(1.521138276 
    - 5.066619934 \, l_t 
    - 3.256800825 \, l_t^2 
    - 0.5625 \, l_t^3 \bigg) 
    \nonumber \\ &&
    \hspace{0.5cm} {} 
    +  C_F^2 \,C_A\,
    \bigg(1.236336499 
    + 6.680264837 \,l_t 
    - 0.1895515129 \, l_t^2 
    - 1.031245 \, l_t^3 \bigg)
    \nonumber \\ &&
    \hspace{0.5cm} {}{} 
    + C_F^2 \, T_F\, \bigg(-4.596159345  
    - 2.353230612 \,l_t 
    + 0.3587005501 \,l_t^2 
    + 0.375  \,l_t^3 
    \nonumber \\ &&
    \hspace{1.0cm} {}
    + n_l\, \bigg(-2.313187377 
    - 0.9994272257 \,l_t 
    + 0.3587005501 \,l_t^2 
    + 0.375 \,l_t^3 \bigg)\bigg)
    \nonumber \\ &&
    \hspace{0.5cm} {} 
    + C_F\, C_A^2 \, \bigg(0.7437683464 
    - 3.881114283 \,l_t 
    + 2.26189568 \,l_t^2 
    - 0.4201388889 \,l_t^3 \bigg)
    \nonumber \\ &&
    \hspace{0.5cm} {}
    + C_F\, C_A\,T_F\, \bigg(2.503654176 
    + 1.794781882 \, l_t 
    - 1.279484737 \, l_t^2 
    + 0.3055555556 \,l_t^3 
    \nonumber \\ &&
    \hspace{1.0cm} {}
    + n_l \,\bigg(-1.370511922  
    + 3.449430465 \, l_t 
    - 1.279484737 \, l_t^2 
    + 0.3055555556 \, l_t^3 \bigg) 
    \bigg) 
    \nonumber \\ &&
    \hspace{0.5cm} {}
    + C_F\, T_F^2 \, \bigg(0.6880486468 
    + 0.4672064147 \,l_t 
    + 0.1227851037 \,l_t^2 
    - 0.05555555556 \,l_t^3 
    \nonumber \\ &&
    \hspace{1.0cm} {}  
    + n_l \, \bigg(0.8495320699
    +0.3327224357 \, l_t 
    +0.2455702074 \, l_t^2 
    -0.1111111111 \, l_t^3 \bigg) 
    \nonumber \\ &&
    \hspace{1.0cm} {}  
    + n_l^2 \bigg(0.4681052884 
    - 0.1344839790 \, l_t 
    + 0.1227851037 \, l_t^2 
    - 0.05555555556 \,l_t^3\bigg)
    \bigg)  \Bigg]
  \Bigg\}
  \nonumber
  \eea


\begin{thebibliography}{99}

  \bibitem{Chetyrkin:2006bj}
    K.~G.~Chetyrkin, M.~Faisst, J.~H.~Kuhn, P.~Maierhofer and C.~Sturm,
    %``Four-loop QCD corrections to the rho parameter,''
    arXiv:hep-ph/0605201.
    %%CITATION = HEP-PH 0605201;%%
    
  \bibitem{vanRitbergen:1997va}
    T.~van Ritbergen, J.~A.~M.~Vermaseren and S.~A.~Larin,
    %``The four-loop beta function in quantum chromodynamics,''
    Phys.\ Lett.\ B {\bf 400} (1997) 379.
    %[arXiv:hep-ph/9701390].
    %%CITATION = HEP-PH 9701390;%%
    
  \bibitem{Czakon:2004bu}
    M.~Czakon,
    %``The four-loop QCD beta-function and anomalous dimensions,''
    Nucl.\ Phys.\ B {\bf 710} (2005) 485.
    %[arXiv:hep-ph/0411261].
    %%CITATION = HEP-PH 0411261;%%

  \bibitem{Kajantie:2002wa}
    K.~Kajantie, M.~Laine, K.~Rummukainen and Y.~Schroder,
    %``The pressure of hot QCD up to g**6 ln(1/g),''
    Phys.\ Rev.\ D {\bf 67} (2003) 105008.
    %[arXiv:hep-ph/0211321].
    %%CITATION = HEP-PH 0211321;%%

  \bibitem{Schroder:2005hy}
    Y.~Schroder and M.~Steinhauser,
    %``Four-loop decoupling relations for the strong coupling,''
    JHEP {\bf 0601} (2006) 051.
    %[arXiv:hep-ph/0512058].
    %%CITATION = HEP-PH 0512058;%%

  \bibitem{Chetyrkin:2005ia}
    K.~G.~Chetyrkin, J.~H.~Kuhn and C.~Sturm,
    %``QCD decoupling at four loops,''
    Nucl.\ Phys.\ B {\bf 744} (2006) 121.
    %[arXiv:hep-ph/0512060].
    %%CITATION = HEP-PH 0512060;%%

  \bibitem{Boughezal:2006px}
    R.~Boughezal, M.~Czakon and T.~Schutzmeier,
    %``Charm and bottom quark masses from perturbative QCD,''
    arXiv:hep-ph/0605023.
    %%CITATION = HEP-PH 0605023;%%

  \bibitem{Chetyrkin:2006xg}
    K.~G.~Chetyrkin, J.~H.~Kuhn and C.~Sturm,
    %``Four-loop moments of the heavy quark vacuum polarization function in
    %perturbative QCD,''
    arXiv:hep-ph/0604234.
    %%CITATION = HEP-PH 0604234;%%

  \bibitem{Schroder:2005db}
    Y.~Schroder and M.~Steinhauser,
    %``Four-loop singlet contribution to the rho parameter,''
    Phys.\ Lett.\ B {\bf 622} (2005) 124.
    %[arXiv:hep-ph/0504055].
    %%CITATION = HEP-PH 0504055;%%

  \bibitem{Chetyrkin:2006dh}
    K.~G.~Chetyrkin, M.~Faisst, C.~Sturm and M.~Tentyukov,
    %``e-finite basis of master integrals for the integration-by-parts method,''
    Nucl.\ Phys.\ B {\bf 742} (2006) 208.
    %[arXiv:hep-ph/0601165].
    %%CITATION = HEP-PH 0601165;%%

  \bibitem{Laporta:2001dd}
    S.~Laporta,
    %``High-precision calculation of multi-loop Feynman integrals by  difference
    %equations,''
    Int.\ J.\ Mod.\ Phys.\ A {\bf 15} (2000) 5087.
    %[arXiv:hep-ph/0102033].
    %%CITATION = HEP-PH 0102033;%%

  \bibitem{Laporta:2002pg}
    S.~Laporta,
    %``High-precision epsilon expansions of massive four-loop vacuum bubbles,''
    Phys.\ Lett.\ B {\bf 549} (2002) 115.
    %[arXiv:hep-ph/0210336].
    %%CITATION = HEP-PH 0210336;%%
    
  \bibitem{Schroder:2005va}
    Y.~Schroder and A.~Vuorinen,
    %``High-precision epsilon expansions of single-mass-scale four-loop vacuum
    %bubbles,''
    JHEP {\bf 0506} (2005) 051.
    %[arXiv:hep-ph/0503209].
    %%CITATION = HEP-PH 0503209;%%

  \bibitem{Kniehl:2005yc}
    B.~A.~Kniehl and A.~V.~Kotikov,
    %``Calculating four-loop tadpoles with one non-zero mass,''
    arXiv:hep-ph/0508238.
    %%CITATION = HEP-PH 0508238;%%

  \bibitem{Veltman:1977kh}
    M.~J.~G.~Veltman,
    %``Limit On Mass Differences In The Weinberg Model,''
    Nucl.\ Phys.\ B {\bf 123} (1977) 89.
    %%CITATION = NUPHA,B123,89;%%

  \bibitem{Awramik:2003rn}
    M.~Awramik, M.~Czakon, A.~Freitas and G.~Weiglein,
    %``Precise prediction for the W-boson mass in the standard model,''
    Phys.\ Rev.\ D {\bf 69} (2004) 053006.
    %[arXiv:hep-ph/0311148].
    %%CITATION = HEP-PH 0311148;%%

  \bibitem{Awramik:2004ge}
    M.~Awramik, M.~Czakon, A.~Freitas and G.~Weiglein,
    %``Complete two-loop electroweak fermionic corrections to
    %sin**2(Theta(lept)(eff)) and indirect determination of the Higgs boson
    %mass,''
    Phys.\ Rev.\ Lett.\  {\bf 93} (2004) 201805.
    %[arXiv:hep-ph/0407317].
    %%CITATION = HEP-PH 0407317;%%

  \bibitem{QCD2L}
    A.~Djouadi and C.~Verzegnassi,
    %``Virtual Very Heavy Top Effects In Lep / Slc Precision Measurements,''
    Phys.\ Lett.\ B {\bf 195}, 265 (1987);
    %%CITATION = PHLTA,B195,265;%%
    A.~Djouadi,
    %``O (Alpha Alpha-S) Vacuum Polarization Functions Of The Standard Model Gauge
    %Bosons,''
    Nuovo Cim.\ A {\bf 100}, 357 (1988);
    %%CITATION = NUCIA,A100,357;%%
    B.~A.~Kniehl, J.~H.~Kuhn and R.~G.~Stuart,
    %``QCD Corrections, Virtual Heavy Quark Effects And Electroweak Precision
    %Measurements,''
    Phys.\ Lett.\ B {\bf 214} (1988) 621.
    %%CITATION = PHLTA,B214,621;%%

  \bibitem{QCD3L}
    L.~Avdeev, J.~Fleischer, S.~Mikhailov and O.~Tarasov,
    %``0 (alpha alpha-s**2) correction to the electroweak rho parameter,''
    Phys.\ Lett.\ B {\bf 336} (1994) 560
    [Erratum-ibid.\ B {\bf 349} (1995) 597];
    %[arXiv:hep-ph/9406363].
    %%CITATION = HEP-PH 9406363;%%
    K.~G.~Chetyrkin, J.~H.~Kuhn and M.~Steinhauser,
    %``Corrections of order O (G(F) M(t)**2 alpha-s**2) to the rho parameter,''
    Phys.\ Lett.\ B {\bf 351} (1995) 331;
    %[arXiv:hep-ph/9502291].
    %%CITATION = HEP-PH 9502291;%%
    K.~G.~Chetyrkin, J.~H.~K\"uhn and M.~Steinhauser,
    %``QCD corrections from top quark to relations between electroweak parameters to order alpha-s**2,''
    Phys.\ Rev.\ Lett.\  {\bf 75}, 3394 (1995).
    %%CITATION = HEP-PH 9504413;%%

  \bibitem{largetop}
    J.~J.~van der Bij and F.~Hoogeveen,
    %``Two Loop Correction To Weak Interaction Parameters Due To A Heavy Fermion
    %Doublet,''
    Nucl.\ Phys.\ B {\bf 283} (1987) 477;
    %%CITATION = NUPHA,B283,477;%%
    R.~Barbieri, M.~Beccaria, P.~Ciafaloni, G.~Curci and A.~Vicere,
    %``Radiative correction effects of a very heavy top,''
    Phys.\ Lett.\ B {\bf 288} (1992) 95
    [Erratum-ibid.\ B {\bf 312} (1993) 511];
    %[arXiv:hep-ph/9205238].
    %%CITATION = HEP-PH 9205238;%%
    R.~Barbieri, M.~Beccaria, P.~Ciafaloni, G.~Curci and A.~Vicere,
    %``Two loop heavy top effects in the Standard Model,''
    Nucl.\ Phys.\ B {\bf 409} (1993) 105;
    %%CITATION = NUPHA,B409,105;%%
    J.~Fleischer, O.~V.~Tarasov and F.~Jegerlehner,
    %``Two loop heavy top corrections to the rho parameter: A Simple formula valid
    %for arbitrary Higgs mass,''
    Phys.\ Lett.\ B {\bf 319} (1993) 249;
    %%CITATION = PHLTA,B319,249;%%
    J.~Fleischer, O.~V.~Tarasov and F.~Jegerlehner,
    %``Two loop large top mass corrections to electroweak parameters: Analytic
    %results valid for arbitrary Higgs mass,''
    Phys.\ Rev.\ D {\bf 51} (1995) 3820;
    %%CITATION = PHRVA,D51,3820;%%
    J.~J.~van der Bij, K.~G.~Chetyrkin, M.~Faisst, G.~Jikia and T.~Seidensticker,
    %``Three-loop leading top mass contributions to the rho parameter,''
    Phys.\ Lett.\ B {\bf 498} (2001) 156;
    %[arXiv:hep-ph/0011373].
    %%CITATION = HEP-PH 0011373;%%
    M.~Faisst, J.~H.~Kuhn, T.~Seidensticker and O.~Veretin,
    %``Three loop top quark contributions to the rho parameter,''
    Nucl.\ Phys.\ B {\bf 665} (2003) 649.
    %[arXiv:hep-ph/0302275].
    %%CITATION = HEP-PH 0302275;%%

  \bibitem{radja}
    R.~Boughezal, J.~B.~Tausk and J.~J.~van der Bij,
    %``Three-loop electroweak correction to the rho parameter in the large  Higgs
    %mass limit,''
    Nucl.\ Phys.\ B {\bf 713} (2005) 278;
    %[arXiv:hep-ph/0410216].
    %%CITATION = HEP-PH 0410216;%%
    R.~Boughezal, J.~B.~Tausk and J.~J.~van der Bij,
    %``Three-loop electroweak corrections to the W-boson mass and
    %sin**2(theta(lept)(eff)) in the large Higgs mass limit,''
    Nucl.\ Phys.\ B {\bf 725} (2005) 3.
    %[arXiv:hep-ph/0504092].
    %%CITATION = HEP-PH 0504092;%%

  \bibitem{Czakon:2006pf}
    M.~Czakon, M.~Awramik and A.~Freitas,
    %``Bosonic corrections to the effective leptonic weak mixing angle at the
    %two-loop level,''
    Nucl.\ Phys.\ Proc.\ Suppl.\  {\bf 157} (2006) 58.
    %[arXiv:hep-ph/0602029].
    %%CITATION = HEP-PH 0602029;%%

  \bibitem{Laporta:1997zy}
    S.~Laporta and E.~Remiddi,
    %``The electron (g(e)-2) and the value of alpha: A check of QED at 1 ppb,''
    Acta Phys.\ Polon.\ B {\bf 28} (1997) 959.
    %%CITATION = APPOA,B28,959;%%

  \bibitem{Bierenbaum:2003ud}
    I.~Bierenbaum and S.~Weinzierl,
    %``The massless two-loop two-point function,''
    Eur.\ Phys.\ J.\ C {\bf 32} (2003) 67.
    %[arXiv:hep-ph/0308311].
    %%CITATION = HEP-PH 0308311;%%

  \bibitem{Kazakov:1983ns}
    D.~I.~Kazakov,
    %``Calculation Of Feynman Integrals By The Method Of 'Uniqueness',''
    Theor.\ Math.\ Phys.\  {\bf 58} (1984) 223
    [Teor.\ Mat.\ Fiz.\  {\bf 58} (1984) 343].
    %%CITATION = TMPHA,58,223;%%

  \bibitem{Bekavac:2005xs}
    S.~Bekavac,
    %``Calculation of massless Feynman integrals using harmonic sums,''
    arXiv:hep-ph/0505174.
    %%CITATION = HEP-PH 0505174;%%

  \bibitem{sectors}
    T.~Binoth and G.~Heinrich,
    %``An automatized algorithm to compute infrared divergent multi-loop
    %integrals,''
    Nucl.\ Phys.\ B {\bf 585}, 741 (2000); 
    %  [arXiv:hep-ph/0004013].
    %%CITATION = HEP-PH 0004013;%%
    %\cite{Binoth:2003ak}
    %\bibitem{Binoth:2003ak}
    T.~Binoth and G.~Heinrich,
    %``Numerical evaluation of multi-loop integrals by sector decomposition,''
    Nucl.\ Phys.\ B {\bf 680}, 375 (2004).
    %[arXiv:hep-ph/0305234].
    %%CITATION = HEP-PH 0305234;%%

  \bibitem{Smirnov:1999gc}
    V.~A.~Smirnov,
    %``Analytical result for dimensionally regularized massless on-shell  double
    %box,''
    Phys.\ Lett.\ B {\bf 460}, 397 (1999).
    %[arXiv:hep-ph/9905323].
    %%CITATION = HEP-PH 9905323;%%

  \bibitem{Tausk:1999vh}
    J.~B.~Tausk,
    %``Non-planar massless two-loop Feynman diagrams with four on-shell legs,''
    Phys.\ Lett.\ B {\bf 469}, 225 (1999).
    %  [arXiv:hep-ph/9909506].
    %%CITATION = HEP-PH 9909506;%%
    
  \bibitem{Czakon:2005rk}
    M.~Czakon,
    %``Automatized analytic continuation of Mellin-Barnes integrals,''
    arXiv:hep-ph/0511200.
    %%CITATION = HEP-PH 0511200;%%

  \bibitem{Anastasiou:2005cb}
    C.~Anastasiou and A.~Daleo,
    %``Numerical evaluation of loop integrals,''
    arXiv:hep-ph/0511176.
    %%CITATION = HEP-PH 0511176;%%

  \bibitem{Melnikov:2000qh}
    K.~G.~Chetyrkin and M.~Steinhauser,
    %``Short distance mass of a heavy quark at order alpha(s)**3,''
    Phys.\ Rev.\ Lett.\  {\bf 83} (1999) 4001;
    %[arXiv:hep-ph/9907509].
    %%CITATION = HEP-PH 9907509;%%
    K.~G.~Chetyrkin and M.~Steinhauser,
    %``The relation between the MS-bar and the on-shell quark mass at order
    %alpha(s)**3,''
    Nucl.\ Phys.\ B {\bf 573} (2000) 617;
    %[arXiv:hep-ph/9911434].
    %%CITATION = HEP-PH 9911434;%%
    K.~Melnikov and T.~v.~Ritbergen,
    %``The three-loop relation between the MS-bar and the pole quark masses,''
    Phys.\ Lett.\ B {\bf 482} (2000) 99.
    %[arXiv:hep-ph/9912391].
    %%CITATION = HEP-PH 9912391;%%

  \end{thebibliography}
\end{document}